\documentclass[superscriptaddress,showkeys,twocolumn,nofootinbib,longbibliography]{revtex4-1}

\usepackage{amsmath}
\usepackage{amssymb}
\usepackage{graphicx}
\usepackage{epsf}
\usepackage{slashed}
\usepackage{enumitem}
\usepackage[czech,english]{babel}
\usepackage[cp1250]{inputenc}
\usepackage{hyperref}
\usepackage{xcolor}
\usepackage{amscd}

\usepackage{cancel}
\usepackage{comment}

\usepackage[only,llbracket,rrbracket,llparenthesis,rrparenthesis]{stmaryrd} 
\usepackage{accsupp} 

\newcommand{\refer}[1]{(\ref{#1})}
\newcommand{\diff}{\mathrm{d}}

\newcommand{\e}{\mathrm{e}}

\usepackage{bbm} 

\usepackage[nodayofweek]{datetime}
\newdateformat{mydate}{\twodigit{\THEDAY}{ }\shortmonthname[\THEMONTH], \THEYEAR}
\usepackage[normalem]{ulem}

\begin{document}

\title{Scattering of kinks in Frankensteinian potentials: Kinks as bubbles of exotic mass and phase transitions in oscillon production.}

\author{Luk\'a\v{s} Rafaj}
\email{lukasrafaj(at)gmail.com}
\affiliation{
Institute of Physics, Silesian University in Opava, Bezru\v{c}ovo n\'am. 1150/13, 746~01 Opava, Czech Republic.
}

\author{Ond\v{r}ej Nicolas Karp\'{i}\v{s}ek}
\email{karponius(at)gmail.com}
\affiliation{
Institute of Physics, Silesian University in Opava, Bezru\v{c}ovo n\'am. 1150/13, 746~01 Opava, Czech Republic.
}

\author{Filip Blaschke}
\email{filip.blaschke(at)physics.slu.cz}
\affiliation{
Research Centre for Theoretical Physics and Astrophysics, Institute of Physics, Silesian University in Opava, Bezru\v{c}ovo n\'am. 1150/13, 746~01 Opava, Czech Republic.
}

\begin{abstract}
We present a dynamical picture of kink-anti-kink scattering in a pair of special, Frankensteinian potentials made of piece-wise quadratic and linear pieces. 
Specifically, we focus on models that support kinks without skin and core regions. We propose an intuitive interpretation for these models as being essentially free massive theories with a built-in particle-pair like production mechanism that enters into the dynamics above certain field-value thresholds. We present results concerning the kink's characteristics depending on these thresholds and the distribution of bouncing windows. We show that the second model exhibits a phase-transition-like property in which the nature of collisions switches from disintegration into a massive wave to production of oscillons for large segments of initial velocities when the field threshold is low enough.

\end{abstract}


\maketitle



\section{Introduction}
\label{sec:I}

Kinks are the simplest topological solitons. They have applications across many disciplines and length-scales, ranging from cosmology as domain-walls \cite{Kibble:1976sj, Vilenkin:1981zs}, defects in crystals and polychains, such as polyacetylene \cite{Su:1979ua}, to undetectably small brane-worlds \cite{ArkaniHamed:1998rs, Antoniadis:1998ig, Randall:1999ee, Randall:1999vf, Rubakov:1983bb}. Kinks also provide a simple, analytic laboratory for exploration of various non-perturbative aspects of quantum field theory \cite{Evslin:2023egm, Evslin:2024wwu, Evslin:2022fzf}.

In this paper, we focus on a relativistic scalar field theory in 1+1 dimensions that supports a kink, namely
\begin{equation}\label{eq:model}
\mathcal{L} = \frac{1}{2}\partial_\mu \phi \partial^\mu \phi - V(\phi)\,.
\end{equation}
To have kinks, all that is needed are two isolated minima (vacua) of the potential $V(\phi)$.
The kink is then a field configuration that interpolates between some pair of such minima at $x = \pm \infty$.
\footnote{
A \emph{kink} is by convention a solution that connects distinct vacua such that $\phi(\infty) > \phi(-\infty)$. The corresponding \emph{anti-kink} connects the same vacua in the opposite order.}
 This is a source of its non-trivial topology and its absolute stability \cite{Manton:2004tk, Shnir:2018yzp}.

The dynamics of kinks, and specifically how they behave during collisions with anti-kinks -- a main focus of this paper -- is governed by the choice of the potential $V$ in a way that is extremely sensitive to details and still not very well understood. 
Indeed, the sine-Gordon (sG) model, i.e., $V_{\rm sG} = 2 \sin^2\bigl(\phi/2\bigr)$ is famously integrable, and the interactions of kinks are completely elastic. On the other hand, the double-well potential $V_{\rm DW} = \bigl(1-\phi^2\bigr)^2/2$ displays phenomena characteristic of non-integrable dynamics of solitons, such as bouncing and production of bions/oscillons.  In their broad sense, these features are universal, meaning that they occur in kink-anti-kink ($K\bar{K}$) scattering for generic potentials. Furthermore, very similar behaviour has been found for other solitons: a striking example is the bouncing and resonant phenomena observed for small oscillons   \cite{Blaschke:2024uec} or Q-balls \cite{Martinez:2025ana} in (1+1)-dimensions. Relatively recently, bouncing has been detected in solitons in higher dimensions as well \cite{Krusch:2024vuy, Bachmaier:2025igf}.

The phenomenon of bouncing was studied in the double-well model since 80ties \cite{Sugiyama:1979mi, Campbell:1983xu, Moshir:1981ja, Belova:1985fg, Anninos:1991un}. Other potentials that came under detailed scrutiny are $\phi^6$ potential \cite{Dorey:2011yw, Adam:2022mmm, Weigel:2013kwa}, the Christ-Lee model \cite{Dorey:2023izf} and $\phi^8$ potential \cite{Gani:2021ttg, Gani:2015cda} among others \cite{Campos:2024ijb, Campos:2025yrx, Goatham:2010dg}.

Typically, $V(\phi)$ is an \emph{analytic} function of the field. However, aside from technical/utilitarian reasons stemming from actual or perceived inconveniences, non-analytic potentials might be as physically relevant as analytic ones.  

Some attention has been paid to cases of potentials that are non-analytic. In particular, the non-existence of the first derivative at the minima gives rise to compact solitons that have been explored in some detail \cite{Arodz:2002yt, Arodz:2007jh, Hahne:2019ela, Hahne:2023dic, Hahne:2022wyl, Streibel:2026yxv}. Such non-differentiable minima might occur due to fixed barriers that are present in the underlying mechanical model of the field theory \cite{Arodz:2002yt} or as a result of a deformation limit of a continuous family of smooth potentials \cite{Hahne:2024qby, Bazeia:2014hja, Bazeia:2019tgt}.

The non-analyticity can, however, be present at other places. Several potentials that display some kind of singularity or non-analyticity either at a local maximum or inflection points were considered \cite{Campos:2019vzf, Karpisek:2024zdj, Belendryasova:2021jgs, Inzunza:2025bbx, Trullinger:1987ie}. 

 \begin{figure*}[htb!]
\begin{center}
\includegraphics[width=0.98\textwidth]{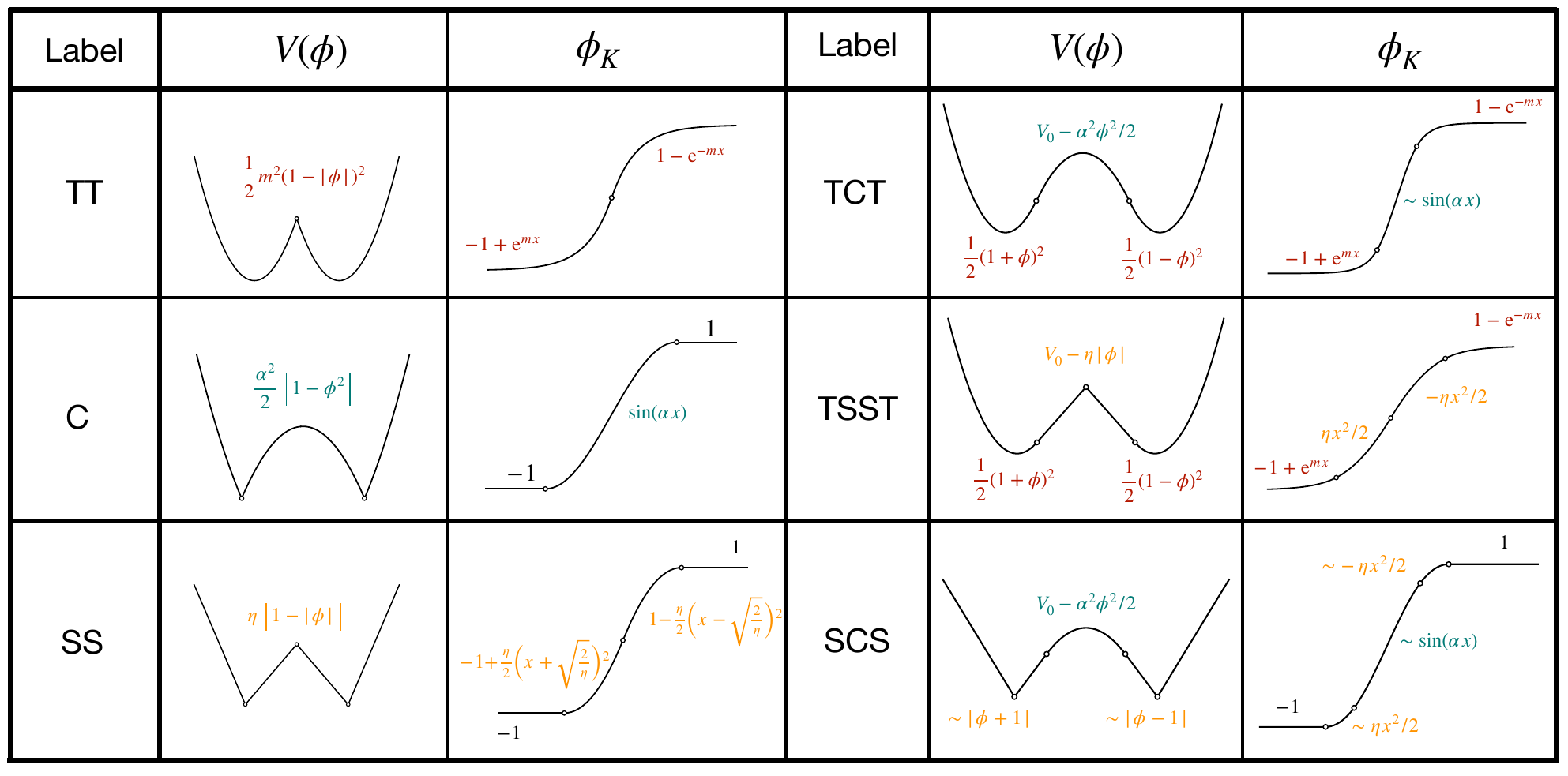}
\end{center}
{
    \caption{\small The simplest Frankensteinian potentials and their corresponding kinks. The vacua are placed at $\pm 1$, and kinks are centered at $x=0$ for simplicity. The labels enumerate the structural pieces of the kinks with T = tail, S = skin, and C = core.}
    \label{fig:simplesttable}
}
\end{figure*}

To isolate the effects of non-analyticity, it is advantageous to make its presence the sole source of non-linearity. To that end, in our previous study \cite{Karpisek:2024zdj}, we introduced a class of potentials made of quadratic (or linear) pieces that are differentiably (or continuously) sewn at certain field values, which we dubbed as \emph{Frankensteinian} (see Fig.~\ref{fig:simplesttable} for the simplest examples). 

As we will stress here, Frankensteinian potentials are essentially free theories with bounds on field values, outside of which they switch to different types of free theories.
This intuition allows us, for instance, to reinterpret a kink as a bound state of particle-like objects that separates regular and exotic (negative $m^2$) Klein-Gordon parts of the potential (see Sec.~\ref{sec:freeish}).

Furthermore, the piece-wise nature of these potentials illuminates the role of the main structural parts of a kink, which we dubbed as \emph{tail}, \emph{skin}, and \emph{core}. These are determined by the local shape of the potential near its minimum (tail), inflection point (skin), and local maximum (core) \cite{Karpisek:2024zdj}.
One can estimate rough extents of these pieces as regions of the field space where the potential is approximately quadratic (near minima and maxima) or linear (near inflection points). However, for smooth potentials, these quantities are necessarily vague. The advantage of Frankensteinian potentials is that they have sharp, non-overlapping boundaries. 

In this paper, we focus on the dynamical properties of two Frankensteinian potentials, namely the tail-core-tail (TCT) and tail-skin-skin-tail (TSST) potential. Both can be regarded as limiting cases of a general symmetric tail-skin-core (TSC) potential, which we describe in App.~\ref{sec:Ia}. 

In Sec.~\ref{sec:II} we present the TCT potential and the main characteristic of its kink, namely the dependence of the Derrick mode and the normal modes on $\beta$. Sec.~\ref{sec:III} serves the same purpose for the TSST potential. We discuss the nature of the observed dynamics through the lens of the pair-production picture in Sec.~\ref{sec:freeish}. The main findings regarding $K\bar{K}$ scattering are collected for both potentials in Sec.~\ref{sec:scattering}. A summary is given in Sec.~\ref{sec:IV}. In App.~\ref{app:A} we comment on our numerical method.

\section{The TCT model}
\label{sec:II}

The `tail-core-tail', or TCT potential is defined as
\begin{equation}
\frac{2V_{\rm TCT}(\phi)}{m^2} = \bigl(1-|\phi|\bigr)^2- \frac{1}{\beta}\theta(\beta - |\phi|)\bigl(\beta-|\phi|\bigr)^2\,,
\end{equation}
 where  $0 < \beta < 1$ is the field value at which the sewing of quadratic functions takes place and where $\theta$ is the Heaviside step function. The TCT potential is the skin-less limit of the general symmetric TSC potential, i.e., $\beta_+\to \beta_- \equiv \beta$ (see App.~\ref{sec:Ia}).
For further use, we also define
\begin{align}
\alpha \equiv & \sqrt{-V_{\rm TCT}^{\prime\prime}(0)} =  m \sqrt{\frac{1}{\beta}-1}\,, \\
 \quad \eta \equiv & V_{\rm TCT}^\prime(\beta) =  m^2\bigl(1-\beta\bigr)\,.
\end{align}

\begin{figure}[htb!]
\begin{center}
\includegraphics[width=0.9\columnwidth]{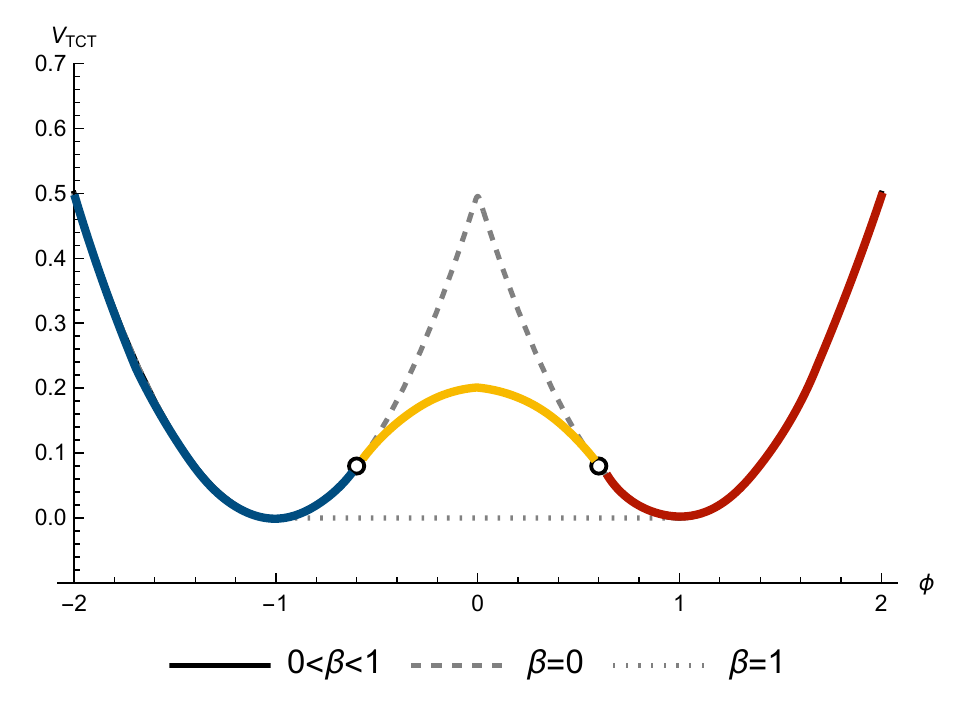}
\end{center}
{
    \caption{\small The TCT potential for a generic value of the sewing point $0<\beta<1$ (solid) and the two limits $\beta =0$ (dashed) and $\beta =1$ (dotted).}
    \label{fig:TCTpot}
}
\end{figure}

The boundary values, i.e., $\beta =0$ or $\beta = 1$, correspond to either trivial or particularly singular dynamics and we cannot expect a smooth continuation beyond this range.
Indeed, the $\beta \to 0$ limit of $V_{\rm TCT}$ potential is $V_{\rm TT}$ potential, which has been studied before \cite{Karpisek:2024zdj} and it only contains trivial dynamics, namely that for all initial velocities $K\bar{K}$ collisions lead only to annihilation into massive waves. On the other hand, the $\beta\to 1$ limit corresponds to the potential
where two quadratic wells are connected by a non-isolated zero inside the interval $\phi \in [-1,1]$ (see Fig.~\ref{fig:TCTpot}). Here, it is impossible to even formulate the kink scattering problem.

We illustrate the $V_{\rm TCT}$ potential and its two limits in Fig.~\ref{fig:TCTpot}.

\subsection{The skin-less kink}

\begin{figure}[htb!]
\begin{center}
\includegraphics[width=0.9\columnwidth]{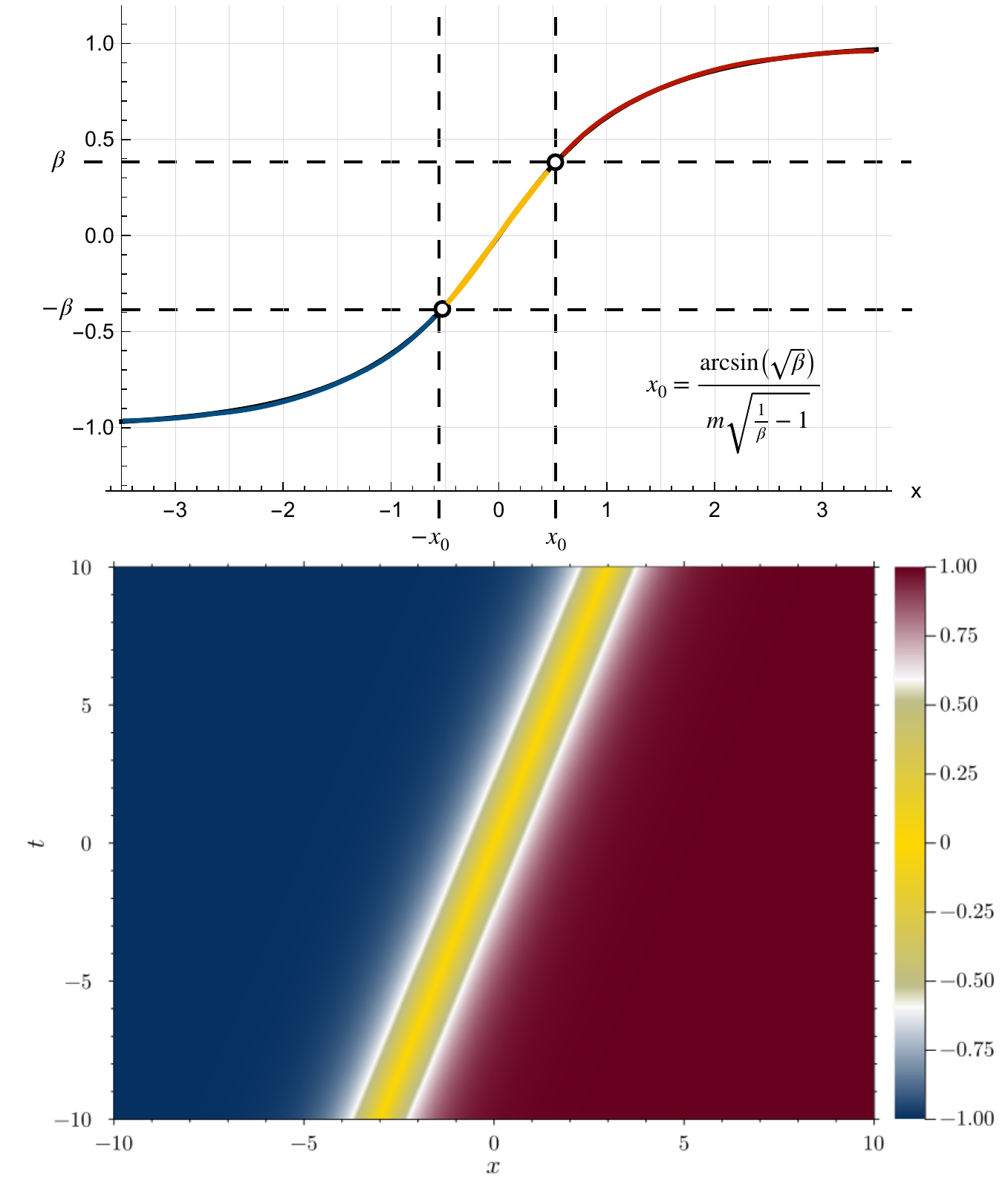}
\end{center}
{
    \caption{\small Upper panel: A TCT kink, centered at origin $x=0$, made of two exponentially decaying tails and a sine core that are glued differentiably at $x = \pm x_0$. Lower panel: Field density plot of a traveling TCT kink, the colors indicate which `type' of field is present. Blue and red represent the Klein-Gordon fields near  -1 and +1 vacua, while gold color stands for the `exotic' negative mass field. Here $\beta = 0.6$.}
    \label{fig:TCTkink}
}
\end{figure}

The TCT kink can be obtained by gluing exponential tails with a sine core (see Fig.~\refer{fig:TCTkink}) and it  is explicitly given as 
\begin{gather}
    \phi_{\rm TCT}  = \theta(-x_0-x)\Bigl(-1+(1-\beta)\e^{m(x+x_0)}\Bigr) \nonumber \\
   +\theta(x-x_0)\Bigl(1-(1-\beta)\e^{-m(x-x_0)}\Bigr)\nonumber \\
     +\bigl(\theta(x+x_0)-\theta(x-x_0)\bigr)\sqrt{\beta}\, \sin\Bigl(m x\sqrt{\tfrac{1}{\beta}-1}\Bigr)\,,
\end{gather}
where the sewing points $\pm x_0$ are determined by the continuity of the first derivative and read
\begin{equation}\label{eq:x0}
    x_0 \equiv \frac{\sqrt{\beta} \arcsin\bigl(\sqrt{\beta}\bigr)}{m \sqrt{1-\beta}} = \frac{\arcsin\bigl(\sqrt{\beta}\bigr)}{\alpha}\,,
\end{equation}
in full accordance to the general formula \refer{eq:sewingpoints1} with $\beta_- = \beta_+ = \beta$.

We can easily integrate the energy density to obtain the mass of the TCT kink as
\begin{align}
    M_{\rm TCT} & =  m \bigl(1-\beta\bigr)^2 + m(1-\beta)(\beta +m x_0) \nonumber \\
    & = 4 V_{\rm TCT}(0) \times R_{\rm TCT}\,,
\end{align}
where in the first line the first term corresponds to the static energy of the tails, while the second term is the core's contribution. In the second line, we introduced the width of the kink as
\begin{equation}
    R_{\rm TCT} = 2 x_0 + 2/m\,,
\end{equation}
which is given by the width of the core $2x_0$ plus the widths of the tails that are given by $1/m$. Again, these formulas are consistent with the general expression \refer{eq:Etot} in the skin-less limit.

\subsection{The Derrick frequency}

We define the Derrick's frequency as the ratio of the kinks mass over the second moment of energy density, i.e.,
\begin{equation}
\omega_D^2 \equiv \frac{M}{Q} = \frac{\int\limits_{-\infty}^{\infty}\diff x\, \bigl(\partial_x \phi\bigr)^2}{\int\limits_{-\infty}^{\infty}\diff x\, x^2\bigl(\partial_x \phi\bigr)^2} \,.
\end{equation}
The integration can be carried out exactly, i.e.,
\begin{equation}
\frac{\omega_{\rm TCT}^2}{m^2} = \frac{6 (\beta -1) \left(m x_0+1\right)}{6 \beta +m x_0 \left(9 \beta +2 (\beta -1) m x_0 \left(m x_0+3\right)-6\right)-3}\,.
\end{equation}
Given the relation \refer{eq:x0}, let us point out that the Derrick frequency over mass squared is a function of  $\beta$ alone. 

\begin{figure}[htb!]
\begin{center}
\includegraphics[width=0.9\columnwidth]{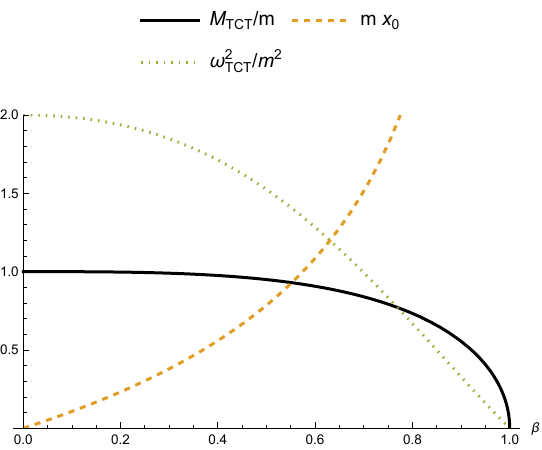}
\end{center}
{
    \caption{\small Dependence of the mass, size of the core and the Derrick frequency on the sewing point $\beta$.}
    \label{fig:allTCT}
}
\end{figure}

In Fig.~\ref{fig:allTCT} we plot how the mass of the kink, the size of the core, and the Derrick frequency change with $\beta$. We can observe that $\beta=0$ correspond to the known TT values, namely $x_0 = 0$, $M_{\rm TT} = m$ and $\omega_{\rm TT}^2 = 2 m^2$. At the opposite extreme, $\beta =1$, the kink solution degenerates to $\phi =0$, hence $M = \omega_{\rm TSST} = 0$ and $x_0 =\infty$. Let us also note that if the Derrick mode is to play any role in $K\bar{K}$ scattering, it should be below the mass threshold, i.e. $\omega_D^2 < m^2$.\footnote{Since the Derrick mode is not a normal mode, it may still play a role in dynamics even if its frequency is above the mass threshold; this makes its crossing not a strict indicator, but rather a useful guide. In our previous study \cite{Karpisek:2024zdj} we observed that this crossing correlated strongly with interesting changes in dynamics. As we will see in later sections, this is not the case in either the TCT or TSST models, and the true utility of the Derrick mode in these models, or Frankensteinian models in general, remains somewhat of an open question.} This happens for $\beta > 0.697$.

\subsection{Normal modes}
\label{subsec:normalTCT}

Following the general discussion of Sec.~\ref{sec:TSCnormmodes}, we find that the effective potential for TCT kink is a  square well with the bottom at $-\alpha^2 = m^2\bigl(1- 1/\beta\bigr)$, while the top of the well rests at $m^2$. 

The solution to the square well Schr\"odinger equation is well known, and the energy spectrum is determined by two transcendental equations for the symmetric and antisymmetric eigenmodes, respectivelly as
    \begin{align}
    \sqrt{\alpha^2 +\omega^2}\tan\bigl(  \sqrt{\alpha^2 +\omega^2} x_0\bigr) & = \sqrt{m^2 -\omega^2}\,,\\
    \sqrt{m^2-\omega^2}\tan\bigl(  \sqrt{\alpha^2 +\omega^2} x_0\bigr) &= -\sqrt{\alpha^2 +\omega^2} \,, 
    \end{align}
which can be also obtained in the skin-less limit, i.e., $\beta_- \to \beta_+ \equiv \beta$, of the general conditions for TSC kink given in Eqs.~\refer{eq:modeseven} and \refer{eq:modesodd}.

To find the minimum value of $\beta$ where the first non-zero frequency mode appears, we plug $\omega^2 = m^2$ for the anti-symmetric case, and we arrive at
\begin{equation}
    \frac{\pi}{2}\sqrt{1-\beta} = \arcsin(\sqrt{\beta}).
\end{equation}
which gives us a numerical value of $\beta \approx 0.646 $.

\begin{figure}
\begin{center}
\includegraphics[width=0.95\columnwidth]{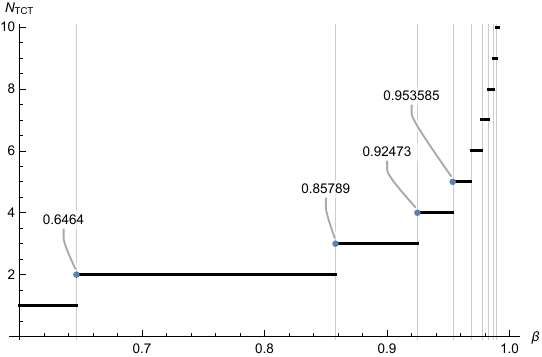}
\end{center}
{
    \caption{\small Number of bound modes for TCT as depending on $\beta$.}
    \label{fig:TCTnoboundmodes}
}
\end{figure}

The number of bound states for any $\beta$ is given by
\begin{equation}
    N  = \left\lceil\frac{2\arcsin(\sqrt{\beta})}{\pi\sqrt{1-\beta}}\right\rceil\,,
\end{equation}
where $\lceil\cdot \rceil$ is the ceiling function. This is shown in Fig.~\ref{fig:TCTnoboundmodes}.

\section{The TSST model}
\label{sec:III}

The `tail-skin-skin-tail', or TSST potential is defined as
\begin{equation}
\frac{2V_{\rm TSST}(\phi)}{m^2} = \bigl(1-|\phi|\bigr)^2-\theta(\beta-|\phi |)\bigl(\beta-|\phi|\bigr)^2\,,
\end{equation}
and corresponds to the limit of the symmetric TSC potential for vanishing core region, i.e. $\beta _- \to 0$ (see App.~\ref{sec:Ia}). Here, again, $ \beta$ represents a field value at which the sewing of linear pieces with the outside quadratic wells takes place, and we restrict $0 < \beta <1$. Note that for $V_{\rm TSST}(\phi)$ we have (c.f. App.~\ref{sec:Ia})
\begin{align}
\alpha \to \infty\,, \quad \eta = m^2(\beta-1)\,.
\end{align}

\begin{figure}[htb!]
\begin{center}
\includegraphics[width=0.9\columnwidth]{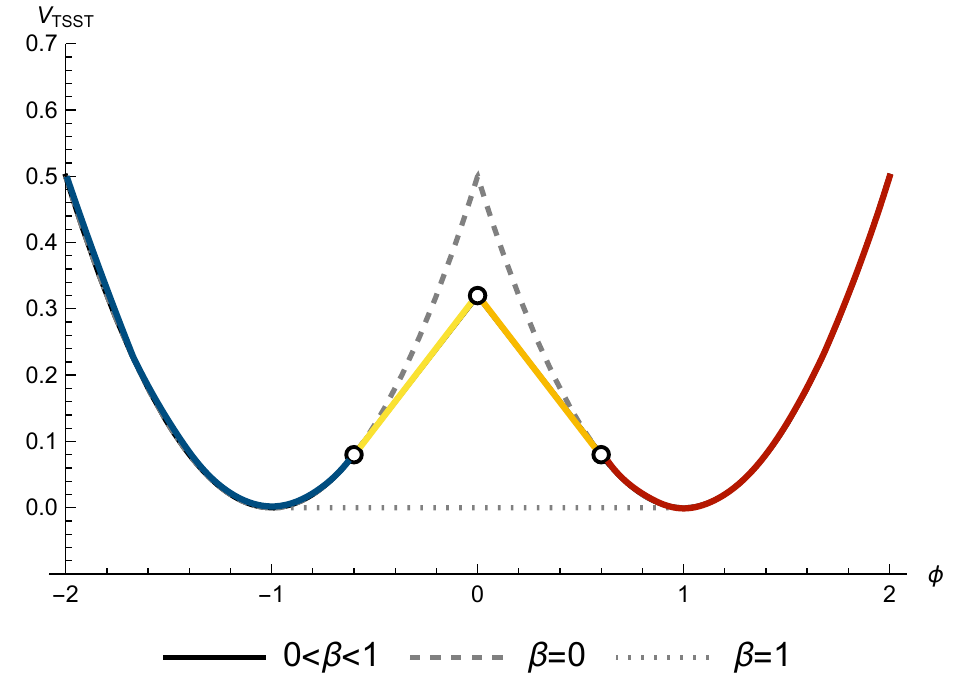}
\end{center}
{
    \caption{\small The TSST potential for a generic value of the sewing point $0<\beta<1$ (black) and the two limits $\beta =0$ (orange, dashed) and $\beta =1$ (green, dotted).}
    \label{fig:TSSTpot}
}
\end{figure}

Similarly to $V_{\rm TCT}$, this potential has the same limiting behaviour, namely $V_{\rm TT}$ as $\beta \to 0$ and $V_{\rm T-T}$ as $\beta \to 1$.
We illustrate the $V_{\rm TSST}$ potential and its two limits in Fig.~\ref{fig:TSSTpot}.

\subsection{The core-less kink}

\begin{figure}[htb!]
\begin{center}
\includegraphics[width=0.9\columnwidth]{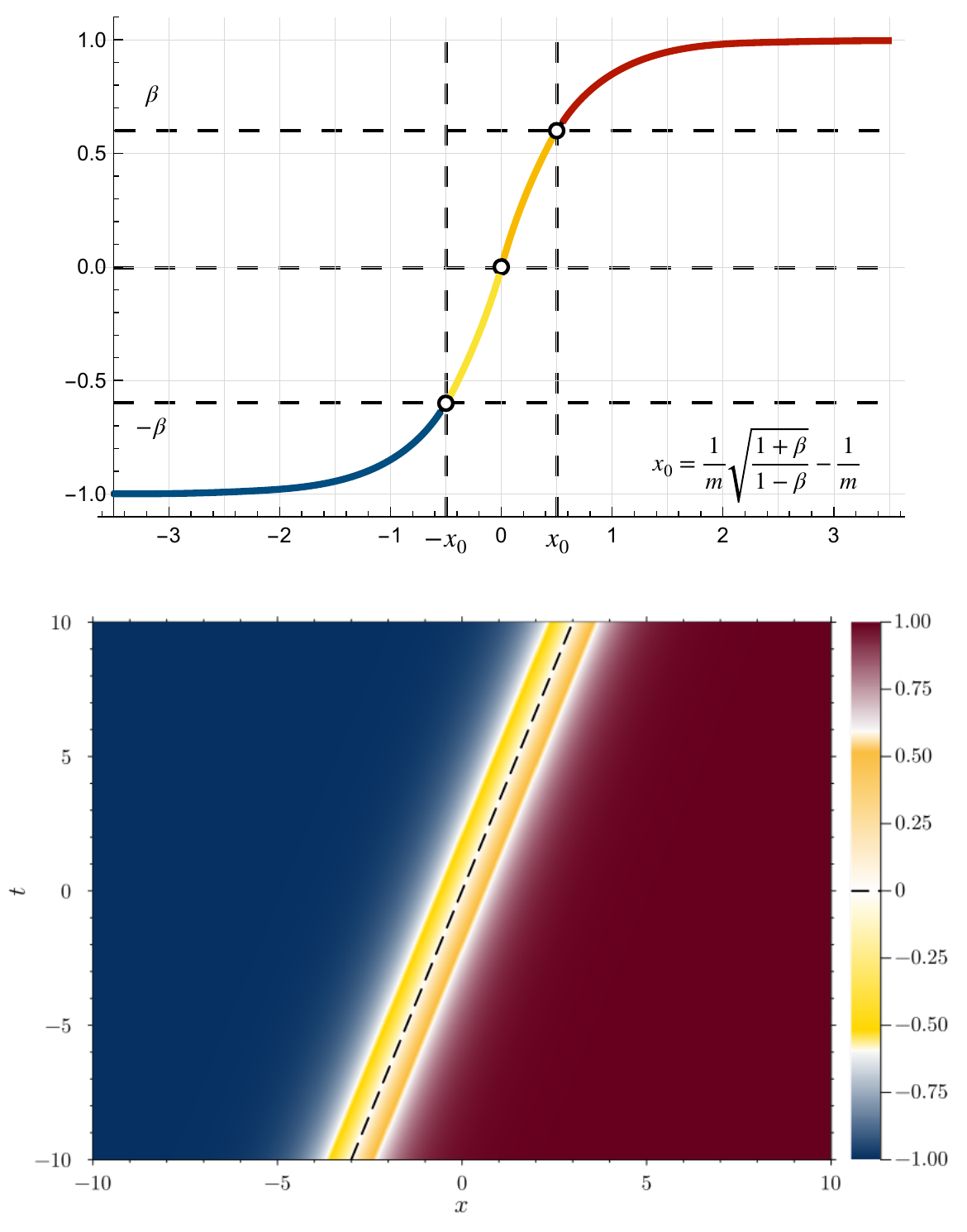}
\end{center}
{
    \caption{\small Upper panel: A TSST kink, here centered at origin $x=0$, made of two exponentially decaying tails and a double quadratic core glued differentiably at $x = \pm x_0$. Lower panel: Field density plot of a traveling TSST kink, the colors indicate which `type' of field is present. Blue and red represent the Klein-Gordon fields near  -1 and +1 vacua, while gold and copper colors stand for the `exotic' linear fields.  Here $\beta = 0.6$.}
    \label{fig:TSSTkink}
}
\end{figure}

The TSST kink is composed of two exponential tails glued to two quadratic skin regions, namely
\begin{gather}
    \phi_{\rm TSST}(x)  = \theta(-x_0-x)\Bigl(-1+(1-\beta)\e^{m(x+x_0)}\Bigr) \nonumber \\
   +\theta(x-x_0)\Bigl(1-(1-\beta)\e^{-m(x-x_0)}\Bigr)\nonumber \\
     +\bigl(\theta(x+x_0)-\theta(x-x_0)\bigr)\frac{m^2}{2}\bigl(1-\beta\bigr)x\bigl(R_{\rm TSST}-|x|\bigr)\,,
\end{gather}
where the sewing points $\pm x_0$ are determined by the continuity of the first derivative and read
\begin{equation}\label{eq:x02}
    x_0 \equiv \frac{1}{m}\biggl(\sqrt{\frac{1+\beta}{1-\beta}}-1\biggr)\,,
\end{equation}
in full accordance with the formula \refer{eq:sewingpoints2} in the limit $\beta_-\to 0$ and with $\beta_+ = \beta$.
Here, again, $x_0$ is well-defined only for $0 <\beta <1$. We depict this kink in Fig.~\ref{fig:TSSTkink}.

We can easily integrate the energy density to obtain the mass of the TSST kink as
\begin{align}
    M_{\rm TSST} & =  \frac{m}{3} \bigl(1-\beta\bigr)^2 +\frac{2}{3}m(1-\beta^2)\sqrt{\frac{1+\beta}{1-\beta}} \nonumber \\
    & =  \frac{1}{3m}V_{\rm TSST}(\beta)+ \frac{4}{3} V_{\rm TSST}(0) \times R_{\rm TSST}\,,
\end{align}
where we used the width of the kink as
\begin{equation}
    R_{\rm TSST} = 2 x_0 + 2/m\,,
\end{equation}
which is given by the width of the core $2x_0$ plus the widths of the tails that are given by $1/m$. Again, these formulas are consistent with the general expression \refer{eq:Etot} in the skin-less limit.

\subsection{The Derrick frequency}

As for the TCT kink, Derrick's frequency and its dependence on $\beta$ can be obtained exactly as
\begin{equation}
\frac{\omega_{\rm TSST}^2}{m^2} = \frac{10 (1-\beta ) (\beta +2 (\beta +1) m x_0+3)}{2 (\beta +1)^2 m x_0-\beta  (5 \beta +2)+15}\,.
\end{equation}
Given the relation \refer{eq:x02}, let us point out that the Derrick frequency over mass squared is a function of  $\beta$ alone. 

\begin{figure}[htb!]
\begin{center}
\includegraphics[width=0.9\columnwidth]{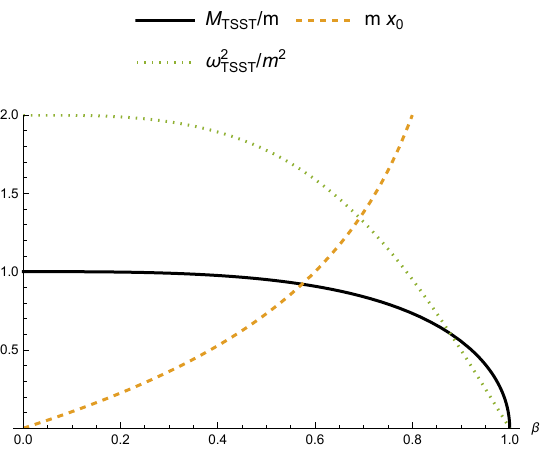}
\end{center}
{
    \caption{\small Dependence of the mass, size of the core, and the Derrick frequency on the sewing point $\beta$ in the TSST model.}
    \label{fig:allTSST}
}
\end{figure}

In Fig.~\ref{fig:allTSST}, we plot how the mass of the kink, the size of the core, and the Derrick frequency change with $\beta$. We can observe that $\beta=0$ correspond to the known TT values, namely $x_0 = 0$, $M_{\rm TT} = m$ and $\omega_{\rm TT}^2 = 2 m^2$. At the opposite extreme, $\beta =1$, the kink solution degenerates to $\phi =0$, hence $M = \omega_D = 0$ and $x_0 =\infty$. For $\beta > 0.788$, Derrick's frequency becomes lower than the mass threshold, i.e., $\omega_{\rm TSST}^2 < m^2$.

\subsection{Normal modes}
\label{subsec:normalTSST}

In contrast with TCT case, normal modes of TSST kink cannot be simply extracted as limiting case of TSC potential studied in Sec.~\ref{sec:TSCnormmodes}, due to singular nature of core-less limit, i.e., $\beta_- \to 0$.
Instead, we find that the effective potential for TSST kink is a square well, with a bottom at 0, while the top of the well rests at $m^2$. More importantly, there is a delta peak in the center $-\lambda \delta (x)$ with
\begin{equation} 
\lambda \equiv 2m\sqrt{\frac{1-\beta}{1+\beta}}\,.
\end{equation}

This delta peak does not have an effect on the odd states since the wave-function is zero there, and we can solve it in the standard way as a finite potential well and get the following transcendental equation
\begin{equation}
\omega \cot (\omega x_0) = -\sqrt{m^2-\omega^2}\,.
\label{eq:odd_tr_equation}
\end{equation}
The number of odd bound states can be given by
\begin{equation}
N_{\mathrm{odd}} = \left \lceil \frac{m x_0}{\pi}-\frac{1}{2}\right \rceil .
\end{equation}

For the even states, the transcendental equation is more complicated and reads
\begin{equation}
\omega \frac{\omega \tan (\omega x_0)+\lambda/2}{\omega -\lambda \tan (\omega x_0)/2} =\sqrt{m^2-\omega^2}\,,
\label{eq:even_tr_equation}
\end{equation}
and the number of even bound states can be well  approximated by
\begin{equation}
N_{\mathrm{even}} \approx \left\lceil \frac{m x_0}{\pi}+\frac{1}{\pi}\mathrm{arctan}\left(\sqrt{\frac{1-\beta}{2\beta}}\right) \right\rceil \,.
\label{eq:number_even_states}
\end{equation}

\begin{figure}
\begin{center}
\includegraphics[width=0.95\columnwidth]{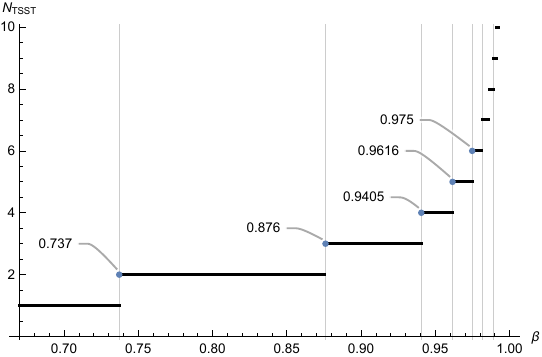}
\end{center}
{
    \caption{\small The dependence of the number of bound modes of the TSST kink on $\beta$.}
    \label{fig:TSSTnoboundmodes}
}
\end{figure}

As we can see from Eq.~\refer{eq:even_tr_equation}, by setting $\omega = 0$, the equation is solved trivially for all $\beta$, which is the consequence of the existence of a zero mode. To find the value of $\beta$ for which the first non-zero frequency bound state appears, we put $\omega = m$ into \refer{eq:odd_tr_equation}, which gives us $\beta = 0.737$.
The number of bound modes depending on $\beta$ is shown in Fig.~\ref{fig:TSSTnoboundmodes}.

\section{Piece-wise Klein-Gordon field as a theory with a built-in Schwinger-like particle pair production.}
\label{sec:freeish}

Due to its piecewise nature, which is at most quadratic in the field, the nontrivial dynamics of a Frankensteinian potential allow for an intriguing re-interpretation. 

The idea is simply to regard these models as free-field theories that have built-in bounds for field values (akin to Schwinger limits) outside of which pair-production-like processes take place. 
Let us illustrate this in the case of the TCT model, although similar reasoning applies to the TSST model or the generic TSC model. 

In the TCT model, we have two types of regimes: i) a regular phase that exits within field values $\phi \in (-\infty, -\beta)$ and $\phi \in (\beta, \infty)$ in which the field obeys usual Klein-Gordon equation $\bigl(\partial^2+m^2\bigr)\phi = 0$ and ii) an ``exotic'' phase lying within the interval $\phi \in [-\beta, \beta]$, where the mass squared parameter is negative, i.e., $\bigl(\partial^2 -\alpha^2\bigr)\phi =0$. 

For finite energy configurations, the exotic phase can only be created in finite pockets (along the $x$ axis). On their own, these pockets would tend to expand; however, at the boundaries $\phi = \pm \beta$, they are pushed back by the pressure of the regular Klein-Gordon fields outside.
 
 It is illustrative to view the boundaries at $\phi^2 = \beta^2$ as particle-like objects that are sources of the exotic field, although these particles do not by themselves contribute to the energy. The intuitive picture is then as follows. Whenever the field enters or exits the interval $\phi \in [-\beta, \beta]$, a pair of these particles is created or destroyed. These particles are not only attracted to each other due to ordinary scalar field interaction, but also repulsed by the mediation of the exotic field in between them. 
These two opposite forces can be exactly balanced, so a static bound-state at a fixed distance is possible. This is nothing but the kink. Since, for a static or boosted kink, the particles either do not move or they move together at uniform velocity  (see Fig.~\ref{fig:TCTkink}), there is no acceleration and hence no radiation. In more general situations, the particles will accelerate towards or away from each other, responding to local changes in the fields, and they will radiate, as seen for instance in Fig.~\ref{fig:TCTScattering}.

In light of this interpretation, a \emph{classical} field-theoretical process, like $K\bar{K}$ collision or oscillon creation and decay, can be viewed -- and understood, at least qualitatively -- in terms that are more in spirit with \emph{quantum} field-theoretical processes in a sense of pair-production.

Indeed, let us consider an \emph{oscillon} and its time-evolution illustrated in late-stage $K\bar{K}$ collision in Fig.~\ref{fig:TCTScattering} c). We can view it as a quasi-periodic pair-production process in which two particles are created, separating the regular $\phi \in (-\infty, -\beta)$ KG phase from the exotic one $\phi \in [-\beta, \beta]$. These pairs, however, do not have sufficient energy to escape their mutual attraction, and they annihilate, but are subsequently produced anew some time later. Each of these "bubbles" releases small energy in the form of escaping radiation, leading to eventual decay of the oscillon. Note that there must be sufficient energy for the Klein-Gordon phase to break into the exotic phase. Hence, we expect that, in this model, the end-stage of the oscillon's lifetime will exhibit a threshold-like, sudden decay into radiation, especially for small values of $\beta$. This should be compared with a gradual decay of small oscillons for smooth potentials.

A different story occurs, when after the creation of the first pair, there is enough energy for the field to break from exotic phase into the other regular phase $\phi \in (\beta, \infty)$ and yet another pair is created -- so that the exotic phase is no-longer a single bubble in the $(t,x)$-plane, but rather a punctured region. These are depicted in the early stages of $K\bar{K}$ collision in Fig.~\ref{fig:TCTScattering} in the panels b) and c).  This is nothing but a \emph{bion} -- a bound state of kink and anti-kink. Note that here, bions are clearly separate entities from oscillons, in contrast to smooth potentials, where their distinction is a much murkier affair.

Hence, in Frankensteinian models supporting kinks, both bions and oscillons will have definite lifetimes that can be quantified as a number of pair-bubbles that were created during their time-evolution. This can further assist in classifying $K\bar{K}$ collisions by the numbers and types of subsequent bubbles (a $K\bar K$ pair is itself a semi-infinite bubble from this point of view). 

This classification may help us uncover deeper intricacies of the dynamic portrait of $K\bar{K}$ scattering.

To that end, in each $K\bar{K}$ collision, we have tracked the number of crossings the field in the center makes with the thresholds, i.e., the number of times $\phi(0,t) = \pm\beta$. The results are presented in Figs.~\ref{fig:TCTKKscan} and \ref{fig:TSSTKKscan}. These plots can be used for finding collisions resulting in short-lived oscillons (small number of $-\beta$ crossings) and collisions resulting in the creation of a $K\bar K$ pair accompanied by a central oscillon (high number of $+\beta$ crossings).  However, in this work, we have not attempted a deeper study of these results in connection with the particle interpretation, and we leave it as future work.

The most important realization, stemming from this interpretation, is the fact that oscillons in Frankensteinian models are of a different type than smooth potentials, namely, they exhibit a sudden decay that is controlled by the field-threshold value $\beta$. A full study into the lifetimes of these objects would be a worthwhile subject of independent research.

For the purposes of this work, let us mention that threshold-like decay of oscillons (and bions) allows us to make a few natural predictions about the $K\bar{K}$ dynamics.

First, for small values of $\beta$, we should observe rather sterile $K\bar{K}$ collisions (= resulting in annihilation), as this implies a high threshold for the pair-production and, hence, oscillon production. As we saw in SubSecs.~\ref{subsec:normalTCT} and \ref{subsec:normalTSST}, normal modes do not appear in both TCT and TSST models until the $\beta$ is relatively large. This implies that for small $\beta$, there is a complete absence of either oscillons or internal models for facilitating a resonant energy transfer mechanism that is necessary for bouncing. 

This prediction is born out by our numerical searches, as can be seen in Figs.~\ref{fig:TCTfreqscan} and \ref{fig:TSSTfreqscan}, where below a certain critical value of $\beta$, no oscillons, bions or $K\bar{K}$ pairs are produced for any initial velocity. In case of the TCT model, this occurs roughly at $\beta_{\rm TCT}^{*} \approx 0.50 $, although the exact transition displays non-trivial velocity dependence. On the other hand, for TSST potential (Fig.~\ref{fig:TSSTfreqscan})), the transition line is almost horizontal for nearly the full interval of initial velocities at roughly $\beta_{\rm TSST}^{*} \approx 0.635$. Thus, in TSST model, we can speak of a \emph{phase transition}, i.e., a sudden switch from annihilation to oscillon production that happens for virtually all initial velocities. However, as Fig.~ \ref{fig:TSSTfreqscan} shows, there is an arcing tendency for very high initial velocities for which the production of oscillons is delayed for higher values of $\beta > \beta_{\rm TSST}^{*} $. This tendency remains somewhat mysterious to us and requires further investigation. 

Secondly, due to the cascade decay of bions (which first turn into oscillons that subsequently decay into massive waves), we should observe a diminished number of higher bouncing windows or their resonant copies with a higher number of internal field oscillations, as both of these characteristic features of $K\bar{K}$ dynamics require long-living bions. It is not hard to guess that long-living bions would be rare in our models, due to the threshold-like decay, in contrast to smooth potentials.
Indeed, this is confirmed in Figs.~\ref{fig:TCTscan} and \ref{fig:TSSTscan}.

\section{Scattering of $K\bar{K}$ pairs in TCT and TSST models}
\label{sec:scattering}

\subsection{TCT model}
\label{subsec:KAKTCT}

Let us now summarize the main features of kink-anti-kink ($K\bar{K}$) scattering for the TCT potential. 

\begin{figure*}
\begin{center}
\includegraphics[width=0.9\textwidth]{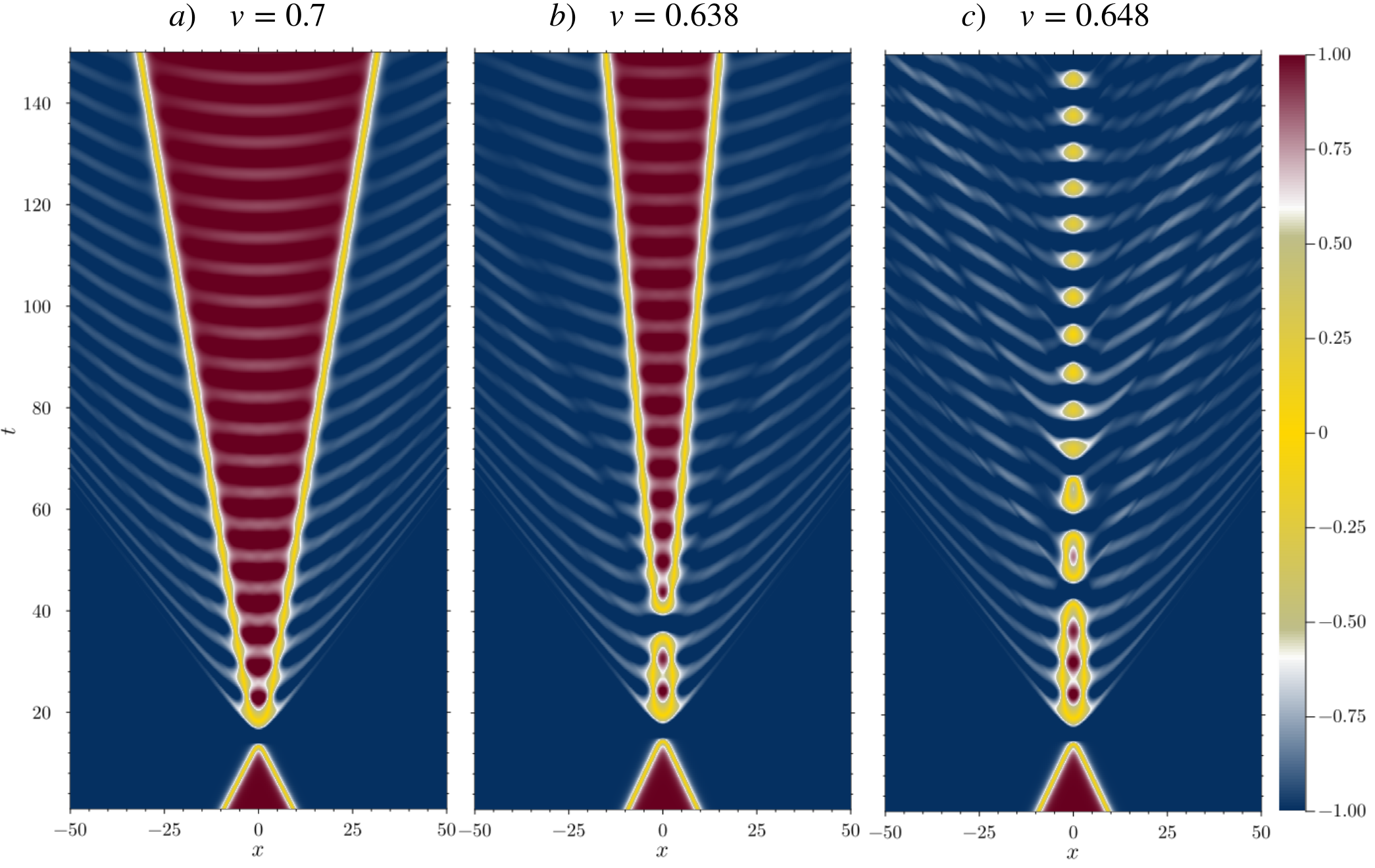}
\end{center}
{
    \caption{\small Examples of $K\bar{K}$ scattering in the TCT model ($\beta = 0.6$) showcasing generic outcomes: a) quasi-elastic collisions, b) bouncing, and c) capture into a bound state and subsequent decay.}
    \label{fig:TCTScattering}
}
\end{figure*}

First of all, we observe all the typical types of outcomes that are present for most symmetric potentials with two vacua, namely quasi-elastic collisions, bouncing, and formation of bions/oscillons as  Fig.~\ref{fig:TCTScattering} illustrates.

\begin{figure*}
\begin{center}
\includegraphics[width=0.99\textwidth]{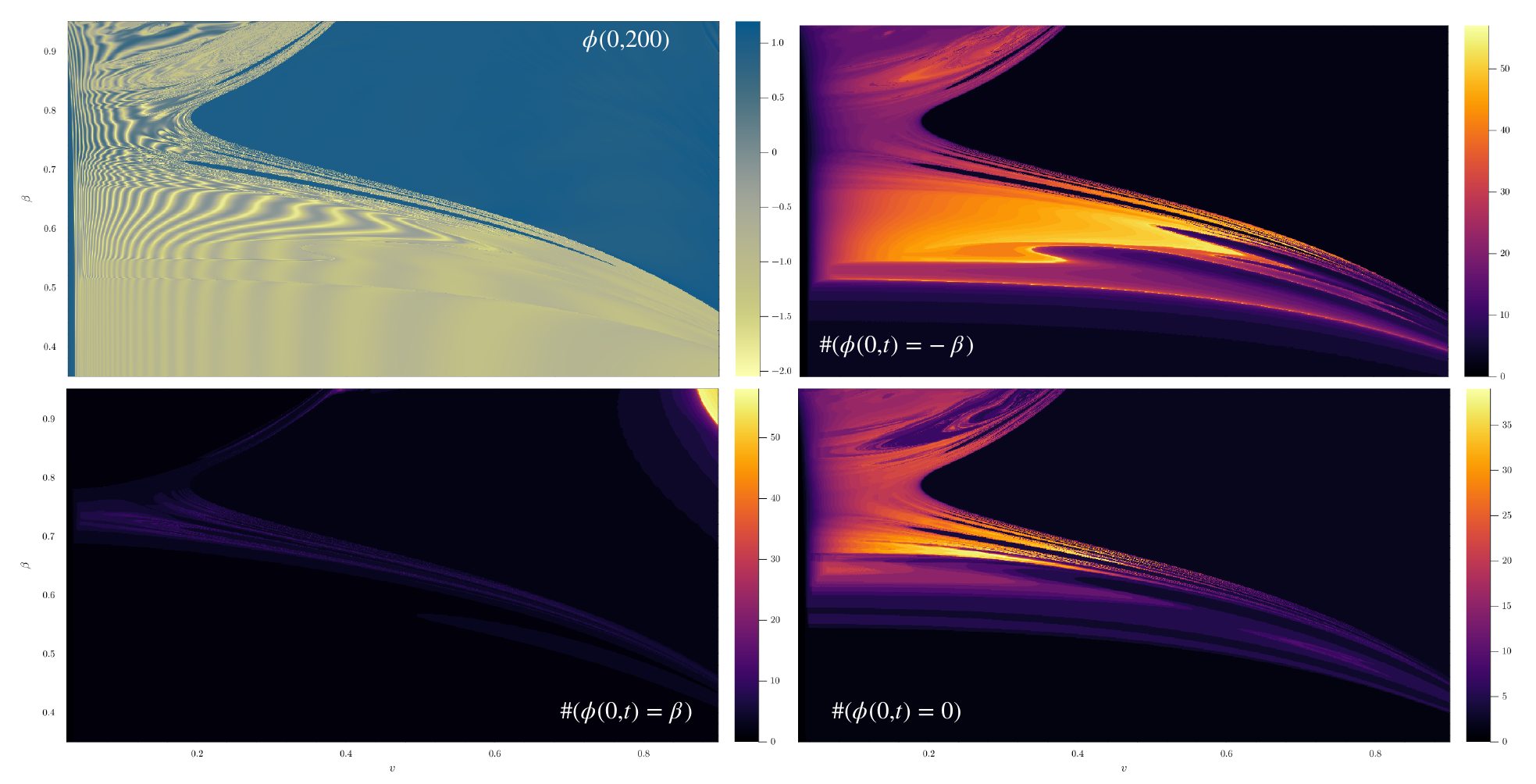}
\end{center}
{
    \caption{\small A scan of $K\bar K$ scattering in the TCT model for a range of initial velocities and $\beta$. Top left: the value of the field at $t=200$. Top right: number of times the field at the center $\phi(0,t)$ crosses the $-\beta$ sewing point. Bottom left: number of times the field at the center $\phi(0,t)$ crosses the $\beta$ sewing point. Bottom right: Number of zeros of the field at the center. }
    \label{fig:TCTKKscan}
}
\end{figure*}

\begin{figure*}
\begin{center}
\includegraphics[width=0.9\textwidth]{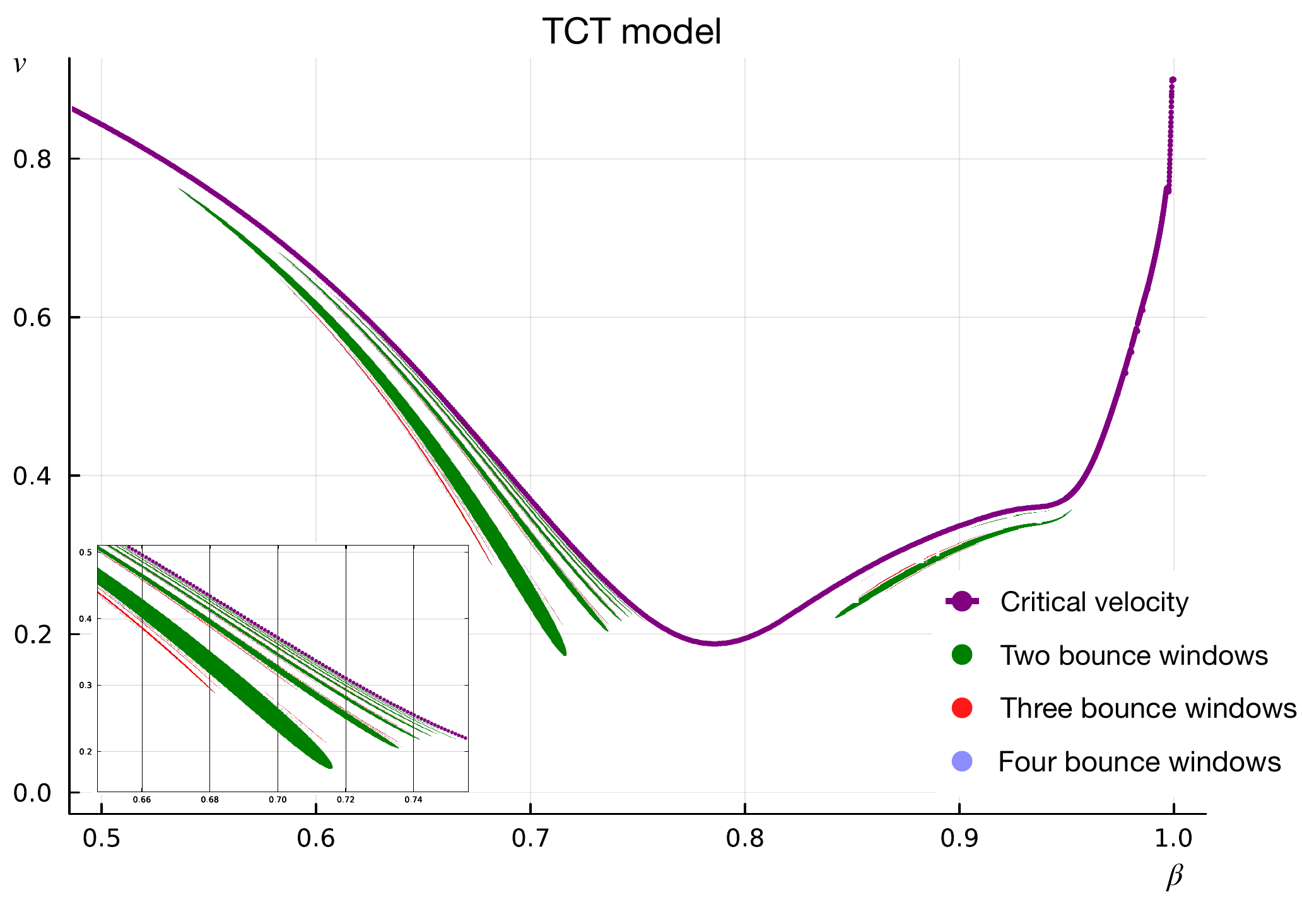}
\end{center}
{
    \caption{\small Dependence of the critical velocity and position of bouncing windows on the sewing point $\beta$ and initial velocity $v$ for the  $K\bar K$ scattering in the TCT model.}
    \label{fig:TCTscan}
}
\end{figure*}

\begin{figure*}
\begin{center}
\includegraphics[width=0.9\textwidth]{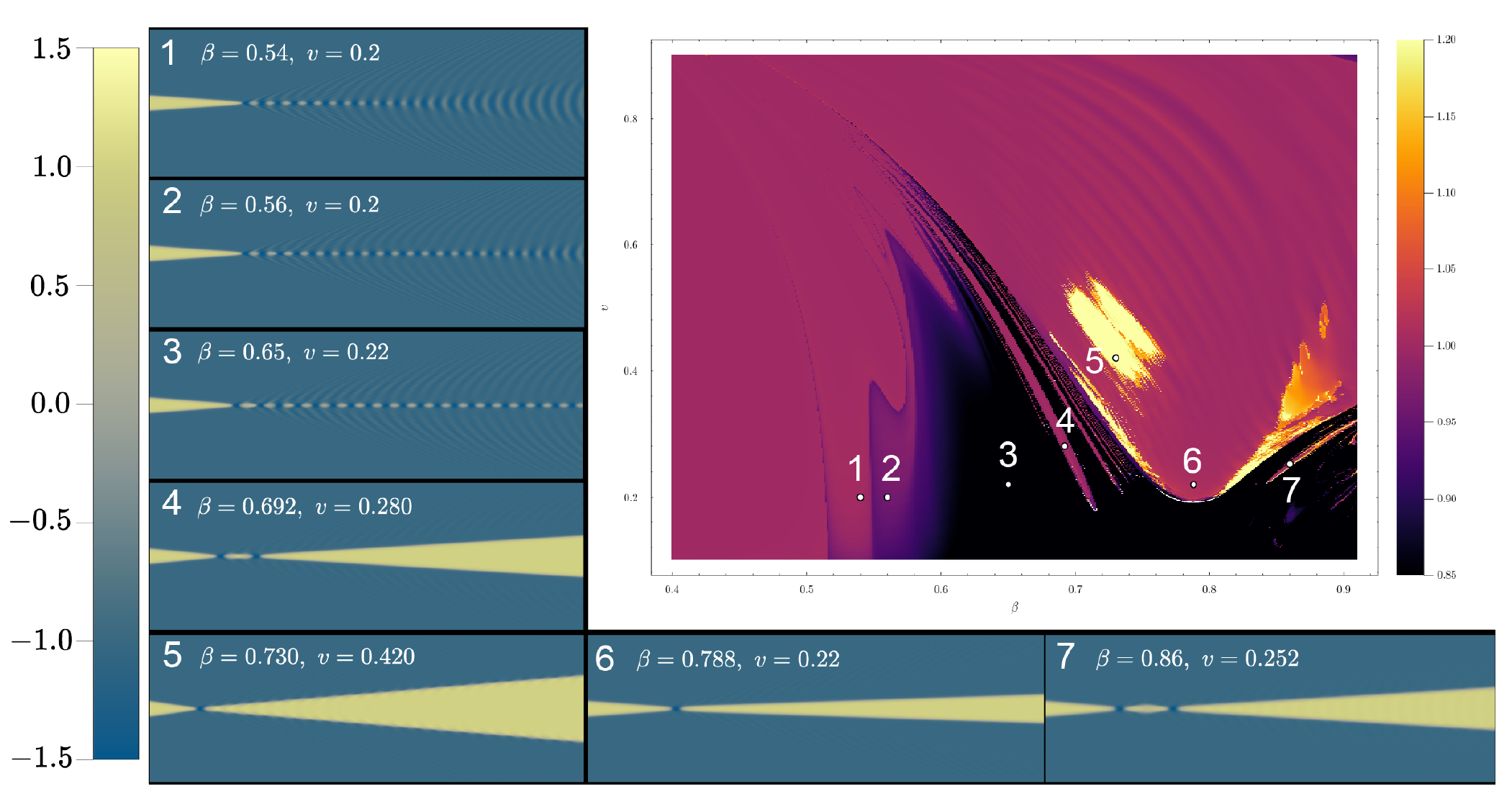}
\end{center}
{
    \caption{\small Dependence of frequency of the field at the center of the collision $\phi(0,t)$ at later stages of the collision, $t=200$~t.u. for the TCT model. We also showcase several collisions at demarcated points.}
    \label{fig:TCTfreqscan}
}
\end{figure*}

The overall dynamical picture of the $K\bar{K}$ scattering is illustrated on Figs.~\ref{fig:TCTKKscan}, \ref{fig:TCTscan}  and \ref{fig:TCTfreqscan}. The first of these displays dependence of the central field, i.e., $\phi(0,t)$, on both the initial velocity and $\beta$ in several ways. In particular, we pay attention to the number of times the field crosses the sewing points, i.e., $\phi = \pm \beta$.\footnote{For completeness, we also track the number of zeros  $\phi = 0$, although they are more important for the TSST potential of the next section, but we keep track of them also here for ease of comparison.} These quantities are particularly useful for identifying the presence of central oscillon in the final stage of the collision. 
 
Indeed, the oscillon cannot exist if the field is solely within a single free regime of the TCT model and must cross either $-\beta$ (if oscillating around -1 vacuum) or $+\beta$ (if oscillating around +1 vacuum) to be long-living. Hence, the areas of the plots on Fig.~\ref{fig:TCTKKscan}, where the number of crossings is large, correspond to $K\bar K$ scatterings for which a central oscillon has formed. 

We see that oscillons (around -1 vacuum) are quite ubiquitous for all $\beta > \beta_{\rm TCT}^{*} \approx 0.5$ and for all ranges of velocities below the critical velocity, as the top right subplot of Fig.~\ref{fig:TCTKKscan} shows.
It is quite reasonable to expect that oscillons are not forming for small $\beta$, as this means that the threshold that needs to be reached is higher, since the sewing points are close to the central hill maximum. Indeed, for $\beta \leq \beta_{\rm TCT}^{*}$ we observe quite sterile scattering where only radiation is produced as the final outcome. 

Fig.~\ref{fig:TCTfreqscan} shows the frequency of the central field at the late stage of the scattering ($t= 200$ t.u.). The presence of an oscillon can be readily seen from the frequency being below the mass threshold, which is fixed at $m = 1$. 

On the other hand, Fig.~\ref{fig:TCTscan} describes the presence (or absence) of bouncing windows in $K\bar{K}$ collisions. The most prominent are two-bounce windows (green) that mainly exist in two strips: for $\beta \in [0.55, 0.75]$ and $\beta \in [0.85,0.95]$. The first strip shows the nested structure of bouncing windows that are accumulating towards the critical velocity curve (violet). Interestingly, while there is a clear beginning of a fractal-like pattern of nesting around the two-bounce windows by three-bounce windows, we could not really find  -- to the level of numerical accuracy -- higher bouncing windows other than very few instances of four-bounce windows.

We do not think this is merely a result of numerical limitations. On the contrary, we believe that in the TCT model, higher bouncing windows are suppressed due to the existence of a sharp threshold. Indeed, higher $K\bar{K}$ bouncing requires the formation of a long-living bion. Its lifetime too exhibits a sudden threshold decay that is even more stringent compared with oscillons (see the discussion in the previous section), and hence, we expect that the number of bounces inherits such a sharp cut-off too.

A similar limitation as on the number of bounces seems to apply to the number of inner oscillations during the bouncing. In other words, the number of recurring two-bounce windows also seems to be finite and small. This is most visible for the green strip towards the right in Fig.~\ref{fig:TCTscan}, where we detected only one (or at most two) two-bounce window(s).

\begin{figure*}
\begin{center}
\includegraphics[width=0.9\textwidth]{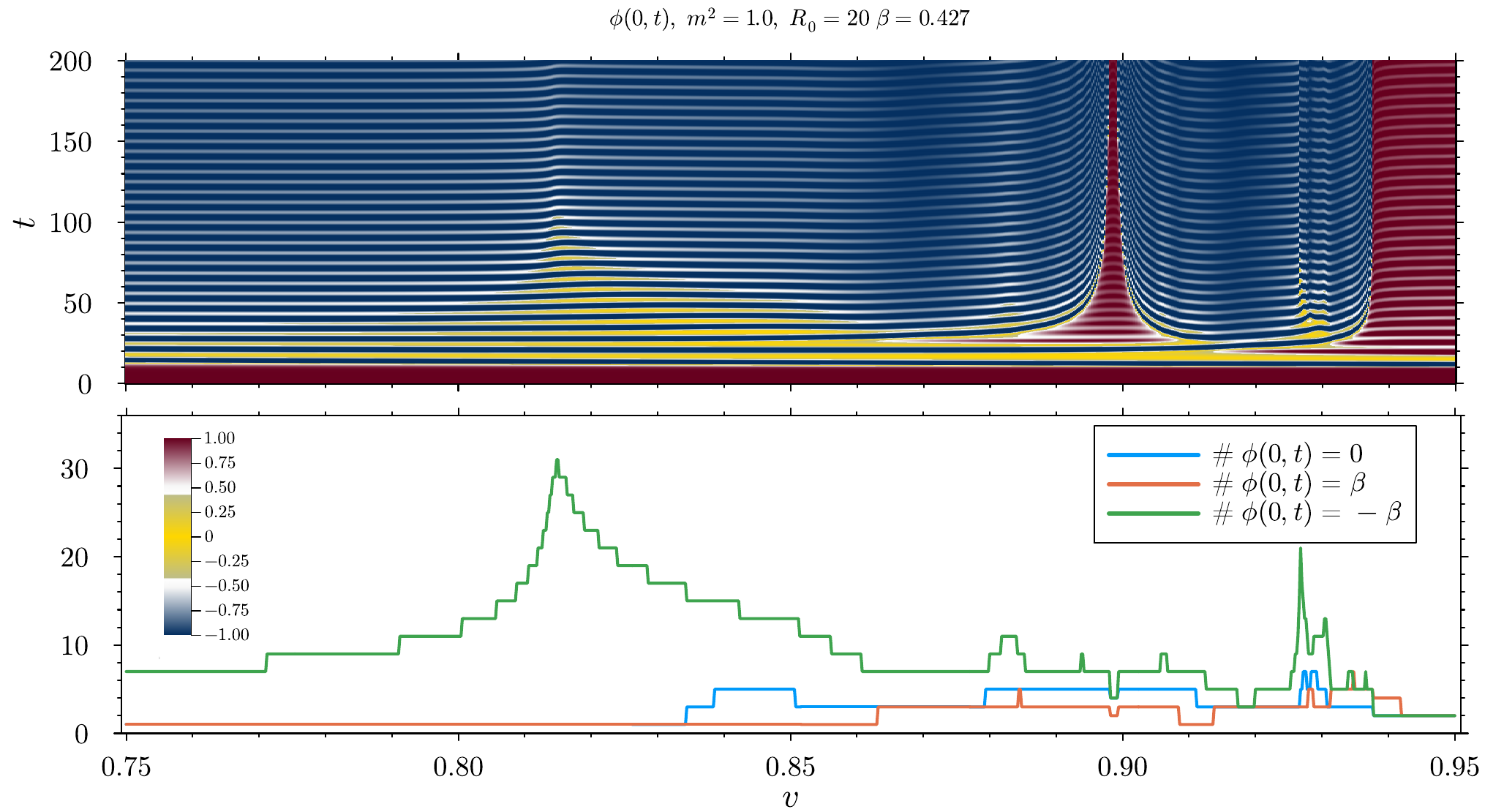}
\end{center}
{
    \caption{\small A velocity map of $K\bar{K}$ scattering in TCT model ($\beta = 0.427$). Top: Field value at the center of collision as dependent on velocity and the first 100 time units. Bottom: Number of crossings of field levels, that is, the number of instances of zeros,  $\phi(0,t) = 0$, and crossing the sewing points, i.e., $\phi(0,t) = \pm \beta$, as dependent on the velocity.}
    \label{fig:TCTmap427}
}
\end{figure*}

We have also find a small two-bounce window that occurs in the vicinity of $\beta \approx 0.4$ for high-end velocities.  We showcase a velocity map in the middle of this range in Fig.~\ref{fig:TCTmap427} for $\beta = 0.427$. 

Let us point out that the first massive mode of the kink occurs at $\beta \geq 0.646$, and the Derrick mode enters the continuum at $\beta \geq 0.697$. Hence, for $\beta = 0.427$ there is no obvious mode for the resonant energy transfer mechanism. It is, therefore, unclear why this spurious bouncing window exists for such low values of $\beta$ (and high values of velocity).

Another anomaly occurs for $\beta$ close to one, namely for $\beta = 0.843$. Again, for high-velocity scattering, a central oscillon is formed, but this time around +1 vacuum, meaning after the separation of the $K\bar K$ pair. These central oscillons are very visible in the number of $+\beta$ crossings on the bottom left panel of Fig.~\ref{fig:TCTKKscan}. 
Let us also point out that the presence of central oscillons for high $K\bar{K}$ velocities was also observed in the hyper-massive models of Ref.~\cite{Hahne:2024qby}.

We will further discuss these results in Sec.~\ref{sec:IV}.

\subsection{TSST model}
\label{subsec:KAKTSST}

\begin{figure*}
\begin{center}
\includegraphics[width=0.9\textwidth]{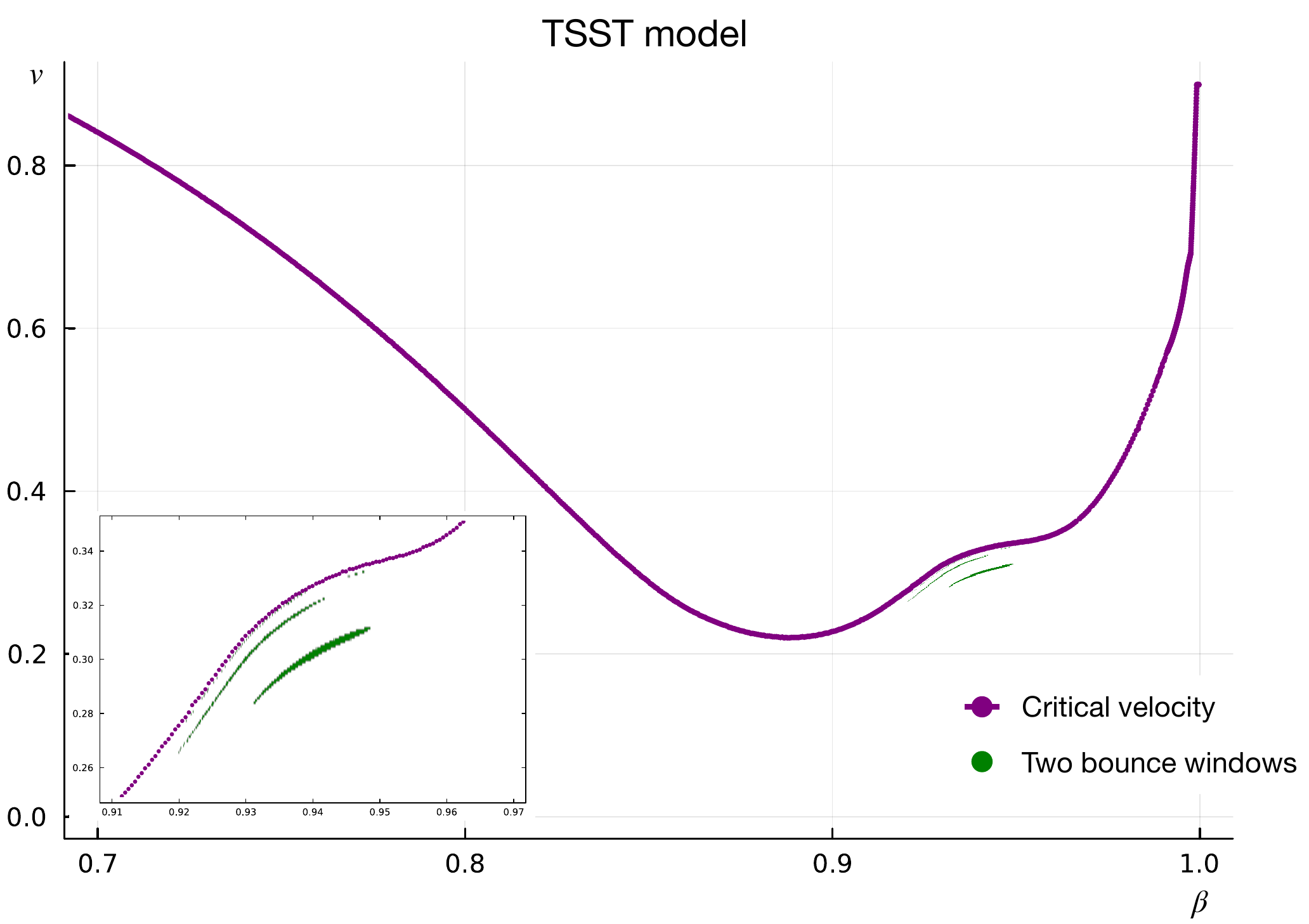}
\end{center}
{
    \caption{\small Dependence of the critical velocity and position of bouncing windows on the sewing point $\beta$ and initial velocity $v$ for the  $K\bar K$ scattering in the TSST model.}
    \label{fig:TSSTscan}
}
\end{figure*}

Now, let us study the $K\bar{K}$ scattering in TSST model. Let us recall that the first massive mode appears for $\beta > 0.737$ and Derrick frequency enters below continuum at $\beta > 0.697$, so below these values one would expect the $K\bar{K}$ phenomenology to be sterile of bounces. Indeed, the TSST model does not support almost any bouncing for any $\beta$, clearly indicating the importance of the core region. To be precise, we have found two instances of two-bounce windows for $\beta \in [0.93, 0.95]$ and no three-bounce windows or higher, see Fig.~\ref{fig:TSSTscan}.  

\begin{figure*}
\begin{center}
\includegraphics[width=0.95\textwidth]{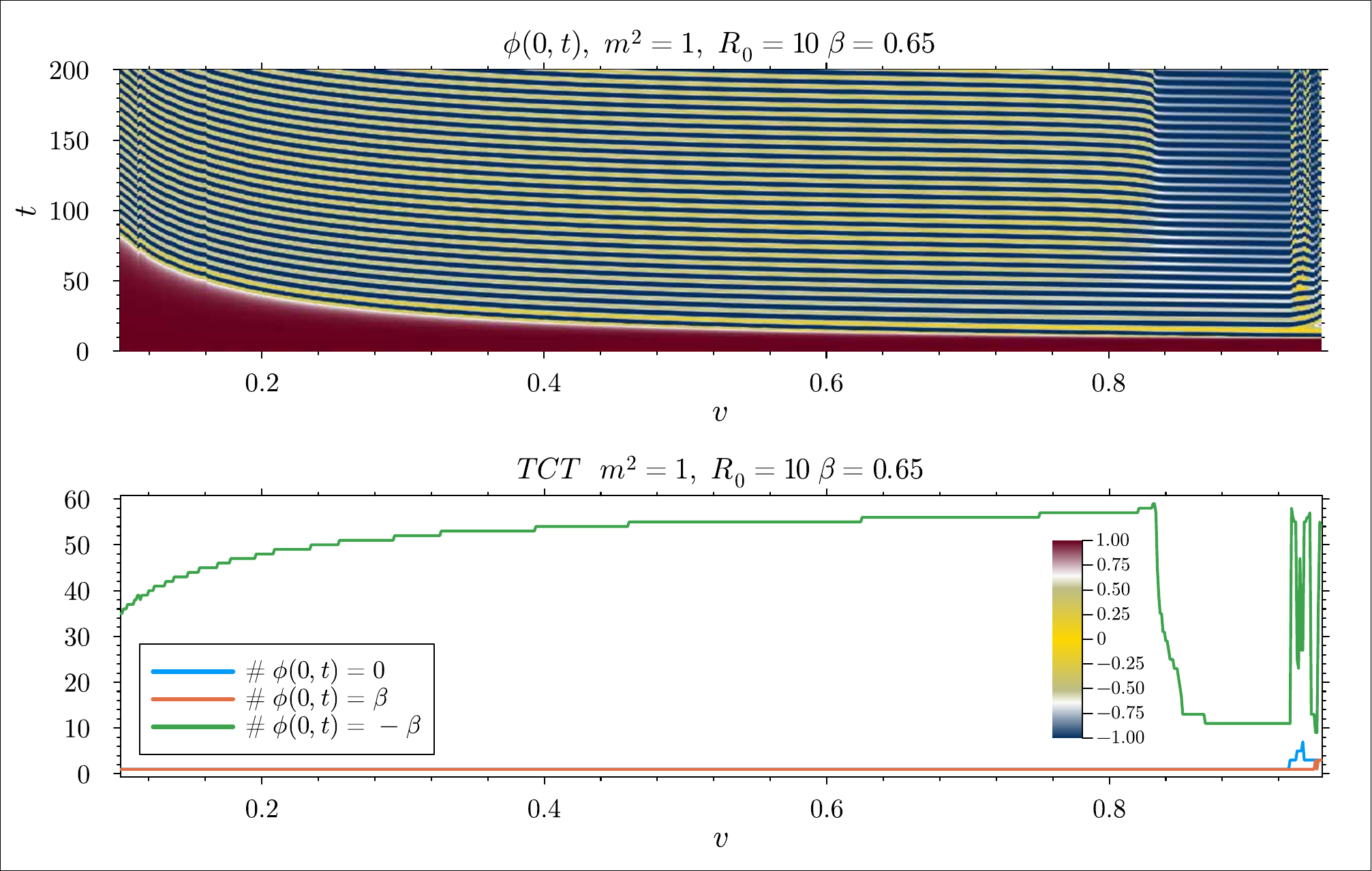}
\end{center}
{
    \caption{\small A velocity map of $K\bar{K}$ scattering in TSST model $(\beta = 0.65)$. Top: Field value at the center of collision as dependent on velocity for $200$ time units. Bottom: Number of times the field value at the center crosses $\phi(0,t) = 0$ (blue), $\phi (0,t) = \beta$ (red), and $\phi (0,t) = -\beta$ (green).}
    \label{fig:TSSTcrossingB065}
}
\end{figure*}

As an example, in Fig.~\ref{fig:TSSTcrossingB065} we showcase a velocity map for $\beta = 0.65$. For this value, the scattering mostly leads to the production of oscillons that, at high velocities, become very short-lived.  

\begin{figure*}
\begin{center}
\includegraphics[width=0.99\textwidth]{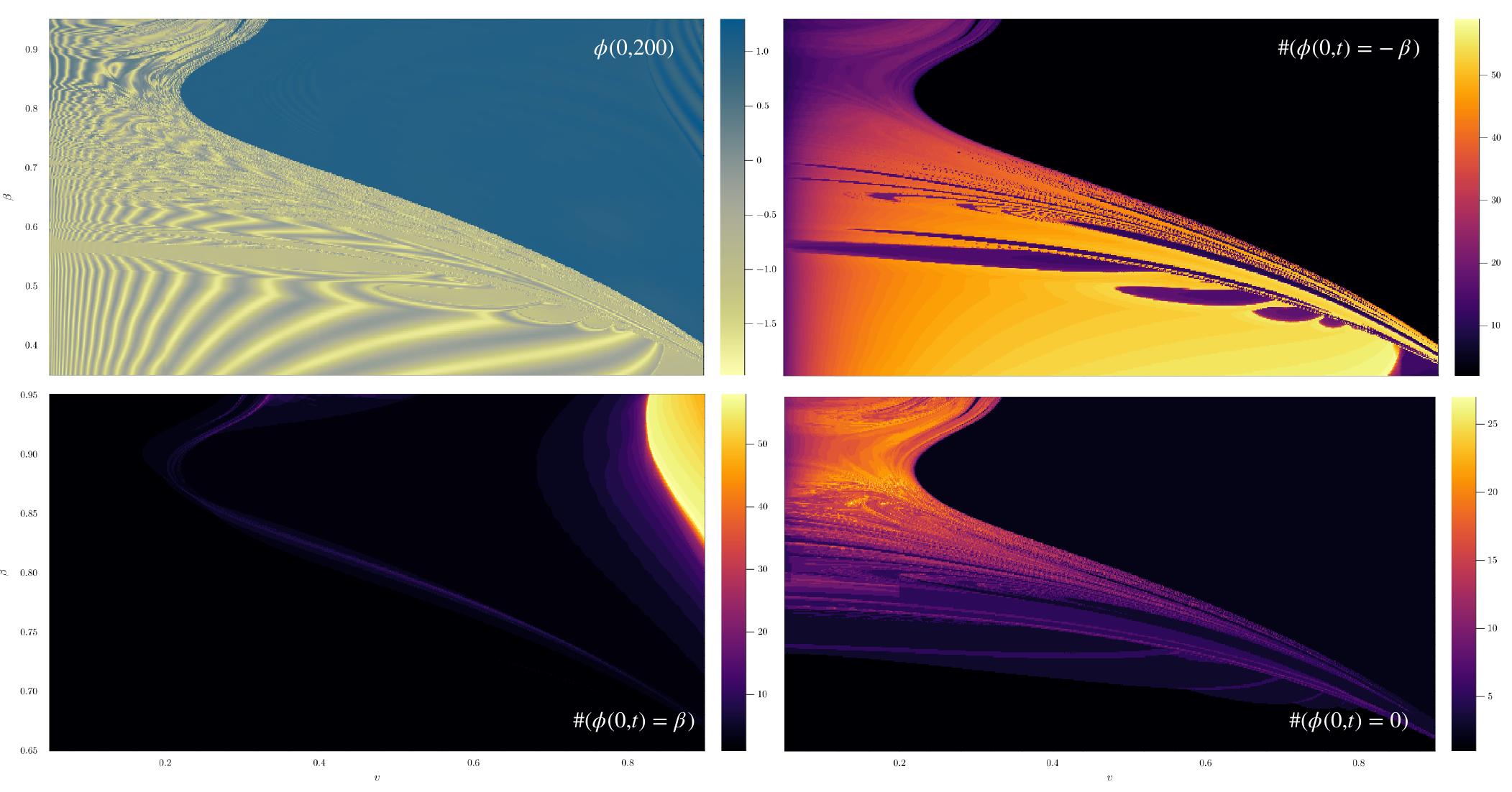}
\end{center}
{
    \caption{\small A scan of $K\bar K$ scattering in the TSST model for a range of initial velocities and $\beta$. Top left: the value of the field at $t=200$. Top right: number of times the field at the center $\phi(0,t)$ crosses the $-\beta$ sewing point. Bottom left: number of times the field at the center $\phi(0,t)$ crosses the $\beta$ sewing point. Bottom right: Number of zeros of the field at the center. }
    \label{fig:TSSTKKscan}
}
\end{figure*}

The full $K\bar{K}$ scattering portrait of the TSST model can be seen in Fig.~\ref{fig:TSSTKKscan}. As noted above,  there are almost no bouncing windows in the TSST model. As we can see from the bottom left picture, there is a region where the number of $+\beta$ crossings is elevated, indicating the existence of a central oscillon on the upper vacuum., as was the case in the TCT model.

The cleanest picture of critical velocity and bouncing windows in the TSST model can be seen in Fig.~\ref{fig:TSSTscan}. We see two significant bouncing windows in the region $\beta \in [0.93, 0.95]$, but we didn't find any higher-order bounding windows. The same comments we made in the TCT model about the lack of higher bouncing windows or copies of two-bounce windows apply here as well.

\begin{figure*}
\begin{center}
\includegraphics[width=0.9\textwidth]{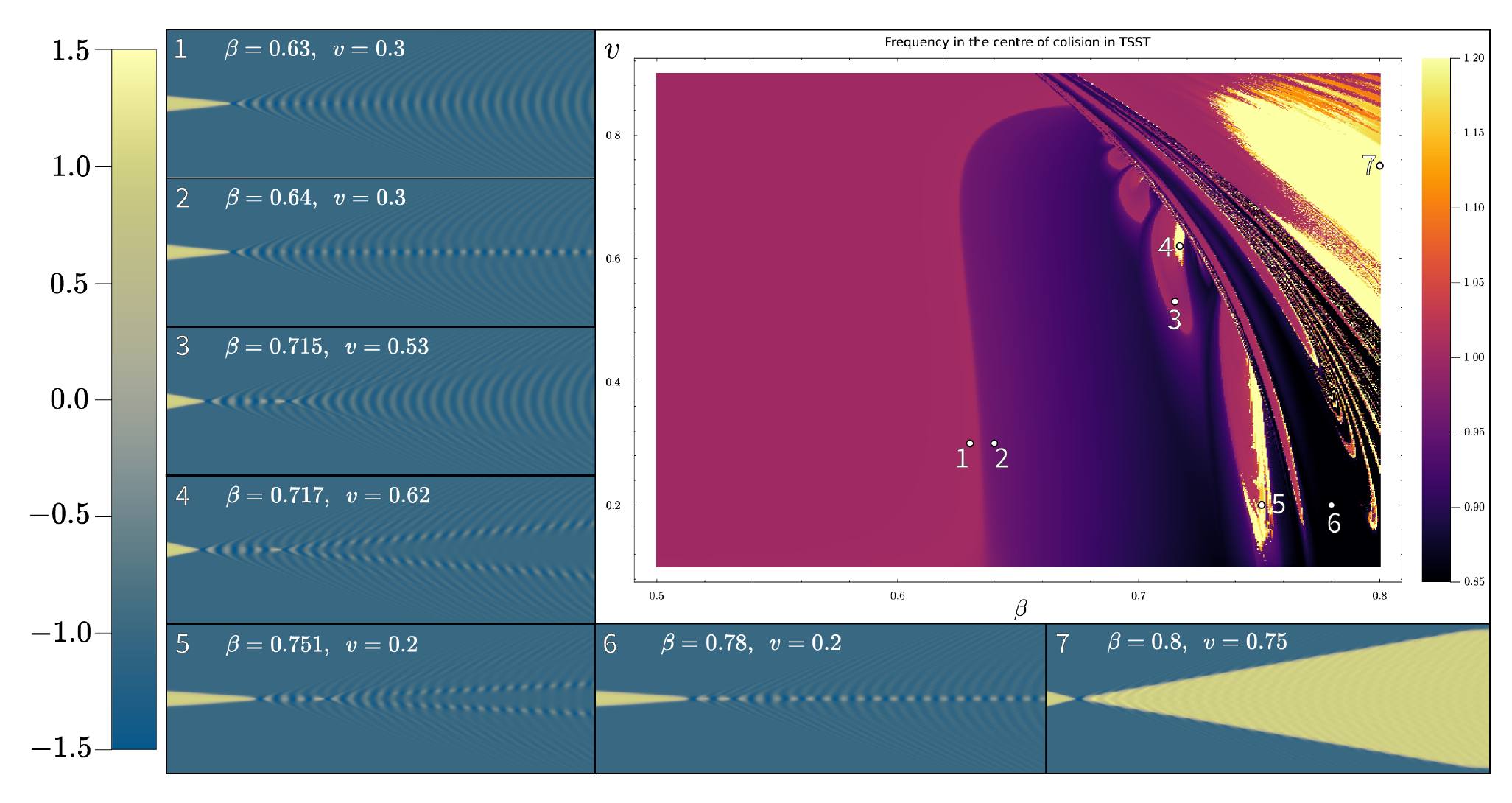}
\end{center}
{
    \caption{\small Dependence of frequency of the field at the center of the collision $\phi(0,t)$ at later stages of the collision, $t=200$~t.u. for the TSST model. We also showcase several  collisions at demarcated points.}
    \label{fig:TSSTfreqscan}
}
\end{figure*}

A good way to detect the existence of oscillons is to simply measure the frequency of the field, as the frequency of the oscillon must be below the mass threshold. In Fig.~\ref{fig:TSSTfreqscan}, we measure the average frequency of the field at the center of collision at sufficiently `post' collision times. The mass threshold in the TSST model with $m = 1$ is $\omega^2 = 1$, and therefore we can associate the final frequencies at or above this value to be massive KG waves, whereas frequencies below this threshold are oscillons. In Fig.~\ref{fig:TSSTfreqscan}, we see a pronounced phase transition from a region where the collision outcomes are just massive waves to a region where oscillons are formed. The precise reason for this phase transition is yet to be understood, but as discussed above, it seems reasonable to associate it with lifetimes of oscillons that depend sensitively on the value of $\beta$.  Aside from a more detailed quantitative understanding of why this transition occurs at $\beta_{\rm TSST}^{*} \approx 0.635$, we also need to explain why the transition to oscillons does not happen for high velocities above $v_{\mathrm{in}} > 0.9$.

 Interestingly, we can also observe frequencies well above the mass threshold, $\omega \approx 1.2$, which appear in the figure as a light-yellow color. Notably, rather than a smooth transition from $\omega = 1$ to $\omega > 1.2$, these frequencies form two sharply distinct regions. This suggests that they correspond to qualitatively different scattering outcomes. Specifically, the dark-red region represents $K \bar{K}$ annihilation into massive waves, whereas the light-yellow region corresponds to the formation of an oscillon pair. These two oscillons subsequently radiate massive waves toward the center, where they interfere. This can be seen in Fig.~\ref{fig:TSSTfreqscan} as example no. $4$.

Let us note that the production of a pair of oscillons is very rarely observed in other models. For instance, it is simply not present in $K\bar{K}$ collisions in the double-well model. It is seen, however, in the hyper-massive models of Ref.~\cite{Hahne:2024qby}.

\section{Discussion}
\label{sec:IV}

In this work, we have investigated kink--anti-kink scattering for a real scalar field under the influence of piece-wise up-to-quadratic potentials, that we dubbed \emph{Frankensteinian potentials}. These models provide a controlled setting in which the structural components of a kink---tail, skin, and core---are sharply separated and can be independently tuned. In particular, we focused on two limiting cases of the symmetric tail--skin--core (TSC) potential: the tail--core--tail (TCT) model, which lacks skin regions, and the tail--skin--skin--tail (TSST) model, which lacks a core.

One of the conceptual outcomes of our study is the reinterpretation of these models as essentially free theories with a built-in particle-like pair-production mechanism. The piece-wise structure introduces sharp field-value boundaries at which the theory effectively switches between free regimes. For instance, in the TCT case, this leads to a picture in which entering the core region corresponds to the creation of a finite ``bubble'' of an exotic negative-mass-squared phase. The interfaces at the sewing points can be viewed as particle-like objects that mediate transitions between ordinary and exotic regimes. This interpretation provides an intuitive understanding of a kink as a stable bubble of exotic phase, and it also sheds light on the dynamical processes observed during the scattering. In particular, it facilitates a clear distinction between oscillons and bions ($K\bar{K}$ bound states), where both can be characterised as quasi-periodic, but ultimately finite sequences of creation and annihilation of particle-pairs akin to quantum-field vacuum fluctuations. 

For both models, we have first made a detailed analysis of static properties of their kinks and how they depend on the sewing parameter $\beta$, including the discussion of normal modes. Let us notice that, although quite distinct in terms of their potential, the TCT and TSST kinks are very similar in these characteristics, as can be seen from Figs.~\ref{fig:allTCT}, \ref{fig:TCTnoboundmodes} and \ref{fig:allTSST}, \ref{fig:TSSTnoboundmodes}.

For both potentials, we have investigated the dynamical portraits of $K\bar{K}$ scattering. Both cases shares many features that sets them apart in terms of $K\bar{K}$ scattering from smooth potentials, due the their piece-wise nature.

Indeed, the most important difference, compared with smooth potentials, is the existence of sharp threshold for creation of oscillons, as the field must penetrate the exotic phase in order to support any long-living structures. This is perhaps most obvious from Figs.~\ref{fig:TCTfreqscan} and \ref{fig:TSSTfreqscan}, where we see that small values of $\beta$ (= high threshold for ``bubble" creation) results in essentially sterile $K\bar{K}$ scattering where the only things produced are massive Klein-Gordon waves. 

In contrast, high $\beta$ regimes makes creation of oscillons very easy. It is then perhaps not so surprising that in both models we see a rare type of outcomes, that is not common for smooth potentials, such as the double-well model. One of such outcome is a formation of a central oscillon on top of the $+1$ vacuum after the separation of a final  $K\bar{K}$ pair. These  are most easily identifiable in the upper-right corners of the bottom-left panels in Figs.~\ref{fig:TCTKKscan} and \ref{fig:TSSTKKscan}, where we display the number of crossings with the upper threshold, i.e., $\# \phi(0,t)= \beta$. It is however, quite puzzling why this happens only for high initial velocities. 
Second type of an unusual production is the creation of an oscillon pair (see examples 3, 4 and 5 in Fig.~\ref{fig:TSSTfreqscan}). Interestingly, double oscillon production is predominantly a feature of TSST model and it is  much less common in TCT model, where it is perhaps suppressed in favor of kink bouncing. 

The bouncing windows shown in Figs.~\ref{fig:TCTscan} and \ref{fig:TSSTscan} clearly demonstrate the most obvious difference between TCT and TSST models. In TCT case, we observe characteristic nested structure of two-bounce windows that accumulates towards the critical velocity, although, as we reported, the number of two-bounce windows with higher number of internal oscillation is limited. Again, this can be attributed to the threshold type nature of oscillons and bions. In contrast, in TSST case, there are almost no bouncing windows, up to two exception in the high-$\beta$ region. It is probable that an especially light central oscillon is responsible for the energy mechanism supporting this bouncing, although is it hard to be quantitative and full explanation would require more detailed analysis. 

There is also an anomalous bouncing window for the TCT potential. In contrast with the above, it occurs for small $\beta$ around  $\beta \approx 0.427$ and lies outside the bound of Fig.~\ref{fig:TCTscan}. Here, however, the bouncing happens at very high velocities and there is no immediate culprit as what facilitates energy-transfer, as the threshold for oscillons is very high and no massive mode yet exist on a kink. It is reasonable to suspect that a dynamical mode may appear in-between kink and anti-kink pair, however a detailed inquiry would have to be made to confirm this.

The observed differences in dynamical portraits of TCT and TSST models do confirm the intuitive thesis that the kink's core seems to be crucial for the phenomena of bouncing, whereas skin seems to be important for  oscillons, in the sense that they dominate the dynamics in the TSST model.  Of course, this can be only taken as a rough heuristic at this point and further investigations are needed. However, taken together, our results demonstrate that the detailed geometric decomposition of a kink -- into tails, skins, and a core -- has direct and observable consequences for $K\bar{K}$ scattering. Frankensteinian models thus provide a convenient laboratory for isolating which structural features are responsible for particular dynamical phenomena.


\acknowledgments

We acknowledge the institutional support of the Research Centre for Theoretical Physics and Astrophysics, Institute of Physics, Silesian University in Opava.
This work has been supported by the grant no. SGS/24/2024 Astrophysical processes in strong gravitational and electromagnetic fields of compact object.
This work has also been partially supported by the KA171/Erasmus+ programme and the internal grant of the Silesian University no IGS/19/2026.

\appendix

\section{General symmetric Frankensteinian model}
\label{sec:Ia}

\begin{figure*}[htb!]
\begin{center}
\includegraphics[width=0.9\linewidth]{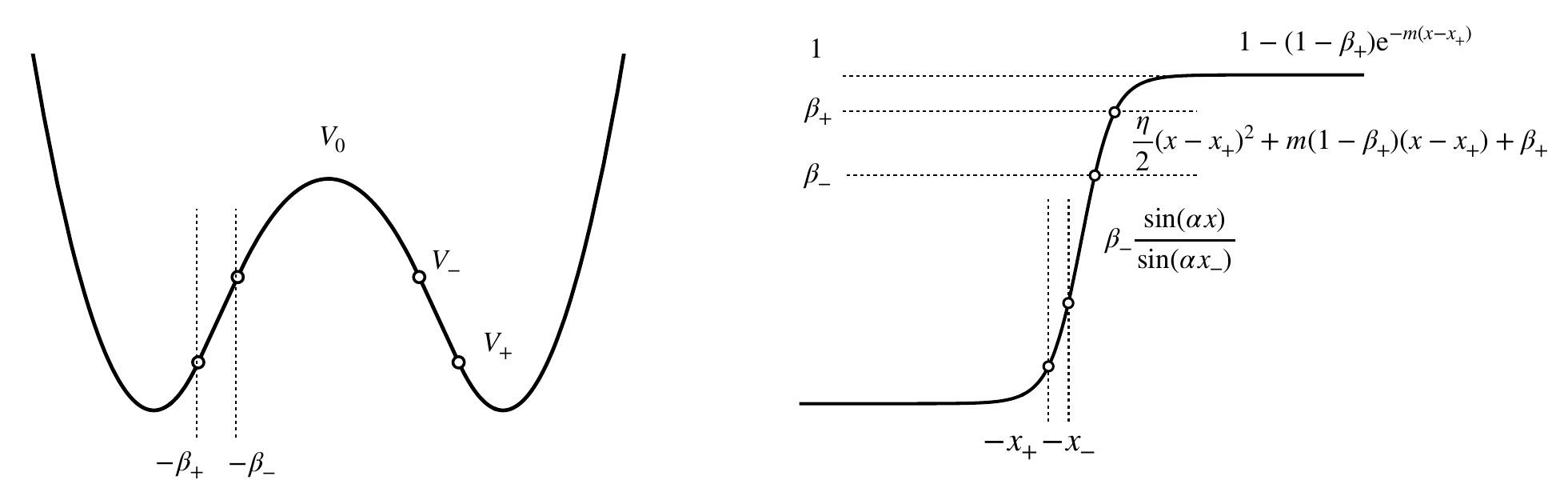}
\end{center}
{
    \caption{\small The geometric characteristics of the symmetric TSC potential and its kink solution.}
    \label{fig:symTSC}
}
\end{figure*}

In this appendix, we describe a generic symmetric Frankensteinian potential consisting of three quadratic and two linear pieces that we denote as (symmetric) TSC potential (see Fig.~\ref{fig:symTSC}, left). Such potential gives rise to a kink that has well-defined tails, core, and skin regions (Fig.~\ref{fig:symTSC}, right).

Without loss of generality, we can always position the two vacua at  $\phi =\pm 1$. 
Given its piece-wise nature together with differentiability across the sewing points, the TSC potential can be fully determined by specifying three numbers. 

A natural choice is to give the positions of the sewing points at $\phi = \beta_{\pm}$ (with the other two mirror images at $\phi = -\beta_{\pm}$) for which we enforce ordering 
\begin{equation}
0< \beta_- < \beta_+ < 1\,.
\end{equation}
The remaining datum can be taken as the perturbative mass, i.e., $m^2 \equiv V_{\rm TSC}^{\prime\prime}(\pm 1)$, which denotes the curvature of the quadratic tail regions.\footnote{In general, we can always change the coordinates to set $m=1$, so there are really only two free parameters, namely  $\beta_{\pm}$. However, we will keep $m$ explicit for convenience.} This allow us to define TSC potential as
\begin{widetext}
{\small \begin{equation}
\frac{2 V_{\rm TSC}(\phi)}{m^2} = \bigl(1-|\phi|\bigr)^2 -\frac{1-\beta_+}{\beta_-}\bigl(\beta_--|\phi|\bigr)^2\theta\bigl(\beta_--|\phi|\bigr)-\bigl(\beta_+-|\phi|\bigr)^2\theta\bigl(\beta_+-|\phi|\bigr)\,,
\end{equation}}
\end{widetext}
where $\theta$ is the Heaviside step function.

At the sewing points, the potential has values $V_{\rm TSC}(\beta_\pm) = V_{\rm TSC}(-\beta_{\pm}) \equiv V_{\pm}$, which are ordered as $ V_- \geq V_+ >0$. Lastly, we denote the height of the central hill as $V_0 \equiv V_{\rm TSC}(0)$.
As a direct consequence of continuity and differentiability we have
\begin{align}
\frac{2V_0}{m^2} = & \bigl(1-\beta_+\bigr)\bigl(1+\beta_+-\beta_-\bigr)\,, \\
\frac{2V_-}{m^2} =&  \bigl(1-\beta_+\bigr) \bigl(1 +\beta_+-2\beta_-\bigr)\,, \\
\frac{2V_+}{m^2}  = &   \bigl(1-\beta_+\bigr)^2\,.
\end{align} 
The triplet $(V_0, V_+, V_-)$ allows an alternative choice of three parameters in defining $V_{\rm TSC}$ compared with the triplet $(m^2, \beta_+, \beta_-)$. 

The main disadvantage of these descriptions is they use parameters that have no direct analogs in smooth potentials. However, there is a unique set of numbers that have analogs for smooth potentials and that also uniquely determine the $V_{\rm TSC}$ potential. Together with $m^2$, they are the slope of the potential at the inflection point, i.e., $\eta \equiv V_{\rm TSC}^\prime(\phi_{\rm inf})$ and the curvature of the top, i.e., $\alpha^2 \equiv  -V_{\rm TSC}^{\prime\prime}(0)$.
These data, i.e., $(m, \eta, \alpha)$, are further advantageous by being invariant under uniform shift of the potential, i.e., $V \to V+c$, as they depend only on its derivatives.

For a TSC potential, we can translate back and forth between these descriptions via relations
\begin{equation}
\beta_- = -\frac{\eta}{\alpha^2}\,, \quad \beta_+ = 1+\frac{\eta}{m^2}\,,
\end{equation}
that are consequences of continuity and differentiability.
Furthermore, let us note the relations
 \begin{align}
  V_+ = & \frac{m^2}{2}\bigl(1-\beta_+\bigr)^2 \equiv \frac{1}{2}W_+^2\,, \\
   V_- = & V_+ -\eta \bigl(\beta_+-\beta_-\bigr) \equiv \frac{1}{2}W_-^2\,, \\
 V_0 = & V_{-} +\frac{\alpha^2}{2}\beta_-^2 \equiv \frac{1}{2}W_0^2\,,
 \end{align}
 where we introduced the quantities $W_{0,\, \pm}$ for future convenience.

\subsection{TSC kink and its properties}

The kink solution consists of two exponential tails, two quadratic skin regions, and a central sine core as illustrated in Fig.~\ref{fig:symTSC} (right).  
The position of sewing points on the $x$-axis is denoted as $x_{\pm}$, and they are determined so that the $\phi_{\rm TSC}(x)$ is differentiable everywhere. This gives us
\begin{align}
x_- =&\, \frac{1}{\alpha}\mbox{arccot}\Bigl(\frac{\alpha}{\sqrt{-\eta}}\sqrt{2+\frac{\eta}{m^2}+\frac{2\eta}{\alpha^2}}\Bigr)\,, \\
x_+ =&\, x_-+\frac{\sqrt{2+\frac{\eta}{m^2}+\frac{2\eta}{\alpha^2}}}{\sqrt{-\eta}}-\frac{1}{m}\,.
\end{align}
Or, alternatively
\begin{align}
 x_- =& \, \frac{\sqrt{2V_0-2V_+}}{2V_0+V_+-V_-}\mbox{arcsin}\Bigl(\sqrt{1-\frac{V_-}{V_0}}\Bigr)\,, \\
 x_+ =& \, x_-+\frac{\sqrt{2V_-}-\sqrt{2V_+}}{2V_0+V_+-V_-}\,,
\end{align}
or
\begin{align}
 \label{eq:sewingpoints1} x_- =& \, \frac{\sqrt{\beta_-}}{m\sqrt{1-\beta_+}}\mbox{arcsin}\Bigl(\sqrt{\frac{\beta_-}{1+\beta_+-\beta_-}}\Bigr)\,, \\
  \label{eq:sewingpoints2} x_+ =& \, x_-+\frac{\sqrt{1+\beta_+-2\beta_-}}{m\sqrt{1-\beta_+}}-\frac{1}{m}\,.
\end{align}

If we now define the extends of various regions as
\begin{equation}
R_{\rm core} = 2x_-\,, \quad R_{\rm skin} = x_+-x_-\,, \quad R_{\rm tail} = 1/m\,,
\end{equation}
where the $R_{\rm tail}$ is chosen as a natural measure of exponentially decaying curve, we can write the BPS mass of the kink, i.e.,
\begin{equation}
E_{\rm TSC} \equiv \int\limits_{-\infty}^{\infty}\diff x\, \bigl(\partial_x \phi_{\rm TSC}\bigr)^2\,,
\end{equation}
 as a sum of contributions from each piece-wise segment, namely:
\begin{equation}\label{eq:Etot}
E_{\rm TSC} = 2E_{\rm tails}+ 2E_{\rm skin}+E_{\rm core}\,,
\end{equation}
where
\begin{align}
\label{eq:Etail2} E_{\rm tail} = & R_{\rm tail} V_+\,, \\
\label{eq:Eskin2} E_{\rm skin} = & \frac{R_{\rm skin}}{3}\bigl(W_+^2+W_-^2 +W_+W_-\bigr)\,, \\
\label{eq:Ecore2} E_{\rm core} = & R_{\rm core} V_0 + \beta_- W_-\,.
\end{align}
Or, equivalently,
\begin{align}
\label{eq:Etail} m E_{\rm tail} = & V_+\,, \\
\label{eq:Eskin} m E_{\rm skin} = & \frac{2}{3}\frac{\sqrt{V_-}^3-\sqrt{V_+}^3}{\sqrt{V_+}}\,, \\
\label{eq:Ecore} m E_{\rm core} = & 2V_0\sqrt{\frac{V_0}{V_+}}\sqrt{1-\frac{V_-}{V_0}}\nonumber \\
 \times & \biggl(\arcsin\Bigl(\sqrt{1-\frac{V_-}{V_0}}\Bigr)+\sqrt{\frac{V_-}{V_0}}\sqrt{1-\frac{V_-}{V_0}}\biggr)\,.
\end{align}

Note that the expression \refer{eq:Etail}-\refer{eq:Ecore} relates the mass of the kink $E_{\rm TSC}$ to a \emph{local} geometric quantities of the potential. 
This should be contrasted with the BPS formula that relates the mass of the kink to an area of the specific curve between the vacua, namely
\begin{equation}\label{eq:BPSmass}
M_K = \int\limits_{-1}^{+1}\diff \phi\, \sqrt{2V(\phi)}\,,
\end{equation}
which is neither local, nor directly related to the potential, but rather to the primitive function of its square root, the superpotential. 

\begin{figure}[htb!]
\begin{center}
\includegraphics[width=0.9\columnwidth]{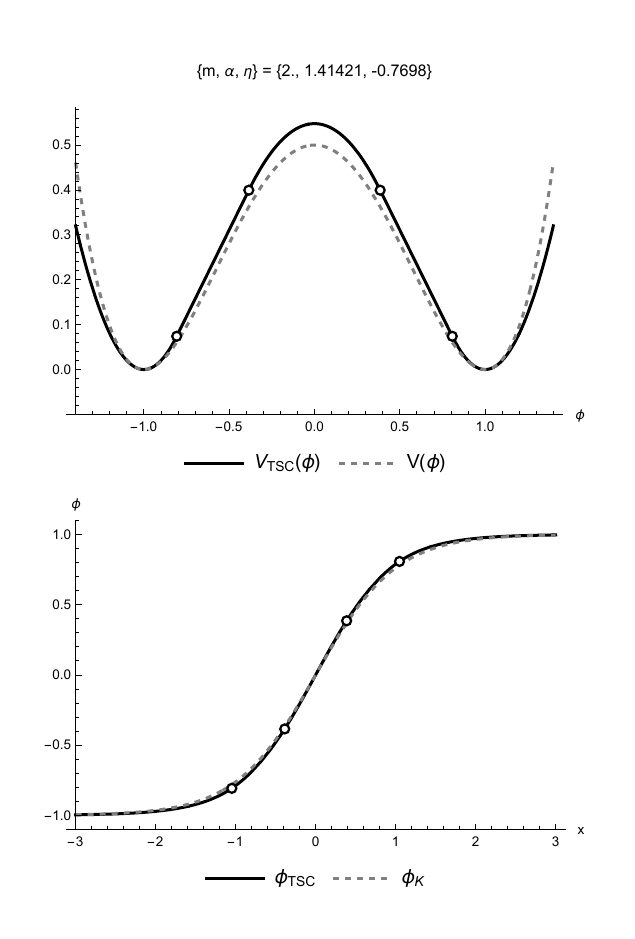}
\end{center}
{
    \caption{\small Comparison between the symmetric TSC potential and its kink with their smooth counterparts in the double well model.}
    \label{fig:compDW}
}
\end{figure}

Let us now -- mostly for curiosity's sake -- compare the TSC kink to some smooth potentials from which we extract the parameters $(m, \alpha, \eta)$. For the double-well model, the parameters read:
\begin{equation}
V = \frac{1}{2}\bigl(1-\phi^2\bigr)^2\, : \quad \{m, \alpha, \eta\} = \{2, \sqrt{2}, -\frac{4}{3\sqrt{3}}\}\,.
\end{equation}
These implies the following values:
\begin{align}
 \{2R_{\rm tail}, 2R_{\rm skin}, R_{\rm core}\} \approx &\ \{1.000, 1.322, 0.774\}\,, \\
 \{2E_{\rm tail}, 2E_{\rm skin}, E_{\rm core}\} \approx &\ \{0.074, 0.569, 0.768\}\,.
\end{align}
We see that the extents of the tails, skins, and the core regions are roughly the same. Interestingly, the contribution to the mass is roughly equal between skin and core, while the contribution of the tails is negligible. Also, the total TSC mass $\approx 1.411$ is not very far from the double well kink value $4/3 \approx 1.334$.\footnote{Let us also note, that the total width $R_{\rm TSC} \approx 3.096$ and the mass $E_{\rm TSC} \approx 1.411$ of the TSC kink are -- perhaps unsurprisingly -- much closer to the so-called mech-kink, for which the corresponding numbers reads $R_M \approx 2.739$ and $E_M \approx 1.461$ \cite{Blaschke:2022fxp}.}
We show the direct comparison in Fig.~\ref{fig:compDW}.

For the sine-Gordon model, which we take rescaled in order to make comparison with the double-well model easier as:
\begin{equation}
V = \frac{2}{\pi^2}\sin^2\Bigl(\frac{\pi}{2}\bigl(1+\phi\bigr)\Bigr) : \hspace{1mm} \{m, \alpha, \eta\} =  \{1, 1, -\frac{1}{\pi}\}\,,
\end{equation}
the respective numbers read
\begin{align}
 \{2R_{\rm tail}, 2R_{\rm skin}, R_{\rm core}\} \approx &\ \{2.000, 1.624, 1.009\}\,, \\
 \{2E_{\rm tail}, 2E_{\rm skin}, E_{\rm core}\} \approx &\ \{0.101, 0.334, 0.402\}\,.
\end{align}
In this case, the numbers do not seem to be markedly tilted in favor of any particular component.
The total energy $\approx 0.838$ is quite close to the actual value $8/\pi^2 \approx 0.811$.

\begin{figure}[htb!]
\begin{center}
\includegraphics[width=0.9\columnwidth]{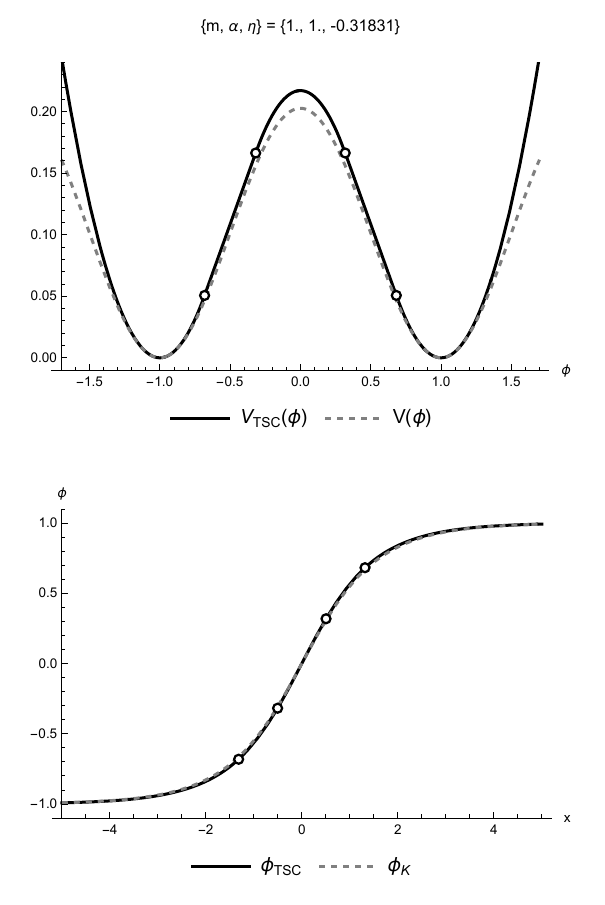}
\end{center}
{
    \caption{\small Comparison between the symmetric TSC potential and its kink with their smooth counterparts in the sine-Gordon model.}
    \label{fig:compSG}
}
\end{figure}

We show the direct comparison in Fig.~\ref{fig:compSG}.

\subsection{Derrick mode}

An important property of the kink is its response to the infinitesimal scaling that is captured by the so-called Derrick's mode. Unlike normal modes, this mode exists independently of the potential and plays an important role in restoring Lorentz covariance in collective coordinate models \cite{Adam:2021gat}. The associated frequency is defined as a ratio of its mass over the second moment of energy density, i.e.,
\begin{equation}
\omega_D^2 \equiv \frac{M}{Q} = \frac{\int\limits_{-\infty}^{\infty}\diff x\, \bigl(\partial_x \phi\bigr)^2}{\int\limits_{-\infty}^{\infty}\diff x\, x^2\bigl(\partial_x \phi\bigr)^2} \,.
\end{equation}
The integrals can be done explicitly. The total mass is given by
\begin{align}
M_{\rm TSC} & = \frac{\eta}{6m}  \bigl(\eta  m (x_+-x_-)^2 \bigl(2 x_-+x_+\bigr)\nonumber \\
& \phantom{=} -6 m x_+-3 \eta  x_-^2+3 \eta  x_+^2-6\bigr)\,,
\end{align}
and, although the second moment can be calculated exactly as well, the formula is too long to display here. 

\begin{figure}[htb!]
\begin{center}
\includegraphics[width=0.9\columnwidth]{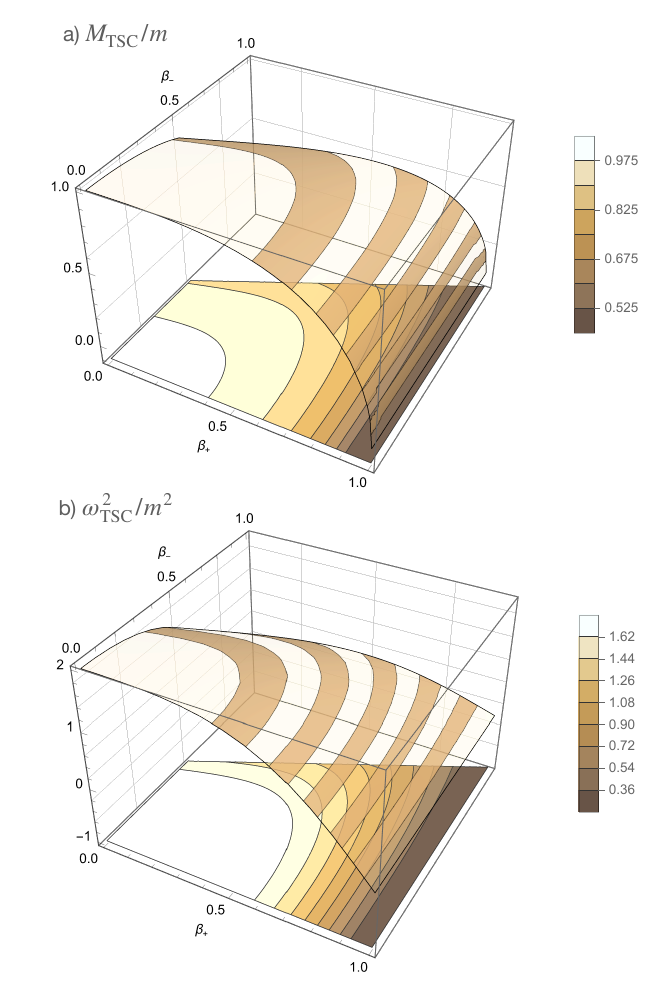}
\end{center}
{
    \caption{\small The dependence of TSC kink on the position of sewing points $\beta_{\pm}$: a) BPS mass, b) Derrick frequency squared.}
    \label{fig:DerrickTSC}
}
\end{figure}

We plot the dependence of the kink's mass and its Derrick frequency in Fig.~\ref{fig:DerrickTSC}

\subsection{Normal modes}
\label{sec:TSCnormmodes}

The normal modes of a kink are obtained by adding to the static solution a small periodic correction, i.e.
\begin{equation}
\phi = \phi_K(x)+ \cos(\omega t)b(x)\,, \quad |b(x)| \ll 1\,,
\end{equation}
and plugging this into the equation of motion, that for a generic scalar field theory in $1+1$ dimensions reads
\begin{equation}
\partial^2 \phi + V^\prime(\phi) = 0\,.
\end{equation}
Retaining only linear terms in $b$, we obtain an effective Schr\"odinger-like eigenproblem
\begin{equation}
-b^{\prime\prime}(x)+ U_{\rm eff}(x)b(x) = \omega^2 b(x)\,,
\end{equation}
where the effective potential is obtained by the shape of the kink solution as
\begin{equation}
U_{\rm eff}(x) = V^{\prime\prime}\bigl(\phi_K(x)\bigr)\,.
\end{equation}

Since the shape of the kink solution is dictated by the shape of the potential in between the vacua, the effective potential is also a product of that shape, although indirectly.
Interestingly, for Frankensteinian potentials that consist of piecewise quadratic or linear functions, the effective potential $U_{\rm eff}$ can be obtained from the potential directly without the need to construct the kink solution first. This is because the second derivatives in each segment are constant and so the effective potential is generically a piece-wise square well.

\begin{figure}[htb!]
\begin{center}
\includegraphics[width=0.9\columnwidth]{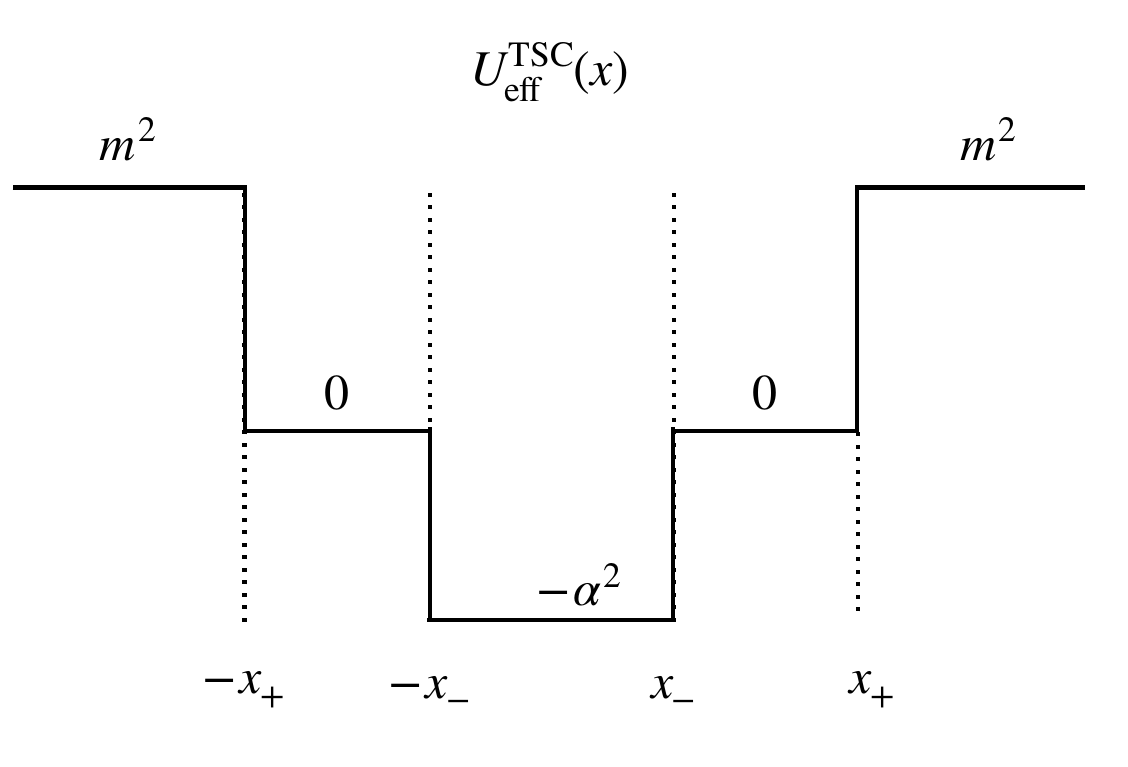}
\end{center}
{
    \caption{\small The effective potential for the TSC model.}
    \label{fig:effpotTSC}
}
\end{figure}

In the TSC case, the effective potential is illustrated in Fig. ~\ref{fig:effpotTSC}.
Since it is symmetric under reflections $x\to -x$, we can discuss normal modes that are either even or odd functions separately. Furthermore, as the kink is stable from topological reasons, there is no danger of tachyonic modes, and we can assume $\omega^2 >0$.

For the even modes, we find the solution of the Schr\"odinger-like eigenproblem to be given in their respective segments as:
\begin{align}
b_{\rm core}^{\rm even}(x) = & \, \frac{-\eta}{\sqrt{\omega^2+\alpha^2}}\frac{\cos\bigl(\sqrt{\omega^2+\alpha^2}x\bigr)}{\sin\bigl(\sqrt{\omega^2+\alpha^2}x_-\bigr)}\,, \\
b_{\rm skin}^{\rm even}(x) = & \, \frac{-\eta}{\omega}\frac{\sin\bigl(\theta+\omega(x_+-|x|)\bigr)}{\cos\bigl(\theta+\omega (x_+-x_-)\bigr)}\,, \\
b_{\rm tail}^{\rm even}(x) = & \, \frac{-\eta}{m}\frac{\exp\bigl(-\sqrt{m^2-\omega^2}(|x|-x_+)\bigr)}{\cos\bigl(\theta+ \omega(x_+-x_-)\bigr)}\,,
\end{align}
where we define the angle $\theta$ as $\omega = m \sin(\theta)$. Since we only look for bounded modes, i.e., $0\leq \omega < m$, we restrict the angle as $0 \leq \theta < \pi/2 $.

Note that these functions agree in their first derivatives at the sewing points $|x| = x_{\pm}$. However, to ensure continuity at the sewing points, the following quantization condition needs to hold:
\begin{multline}\label{eq:modeseven}
\tan\bigl(\theta + m(x_+-x_-)\sin\theta \bigr) \tan\bigl(\sqrt{m^2\sin^2\theta+\alpha^2}x_-\bigr) \\
= \frac{m\sin\theta }{\sqrt{m^2\sin^2\theta+\alpha^2}}\,.
\end{multline}
This condition is fulfilled for $\theta = \omega =0$ and, indeed, the mode function becomes equivalent to $\phi_{\rm TSC}^{\prime}(x)$, which is nothing but the zero mode for the TSC kink.

For the odd modes, we have
\begin{align}
b_{\rm core}^{\rm odd}(x) = & \, \frac{-\eta}{\sqrt{\omega^2+\alpha^2}}\frac{\sin\bigl(\sqrt{\omega^2+\alpha^2}x\bigr)}{\sin\bigl(\sqrt{\omega^2+\alpha^2}x_-\bigr)}\,, \\
b_{\rm skin}^{\rm odd}(x) = & \, \frac{-\eta\, {\rm sgn}(x)}{\sqrt{\omega^2+\alpha^2}}\frac{\sin\bigl(\theta+\omega(x_+-|x|)\bigr)}{\sin\bigl(\theta+\omega (x_+-x_-)\bigr)}\,, \\
b_{\rm tail}^{\rm odd}(x) = & \, \frac{-\eta\, \omega\, {\rm sgn}(x)/m}{\sqrt{\omega^2+\alpha^2}}\frac{\exp\bigl(-\sqrt{m^2-\omega^2}(|x|-x_+)\bigr)}{\sin\bigl(\theta+ \omega(x_+-x_-)\bigr)}\,.
\end{align}
In contrast to even modes, these functions are by construction continuous across the sewing points, however to ensure that the first derivatives agree, we have the condition:
{\small \begin{multline}\label{eq:modesodd}
\frac{\tan\bigl(\theta + m(x_+-x_-)\sin\theta \bigr)}{m\sin\theta}
= -\frac{\tan\bigl(\sqrt{m^2\sin^2\theta+\alpha^2}x_-\bigr) }{\sqrt{m^2\sin^2\theta+\alpha^2}}\,.
\end{multline}}

The quantization conditions \refer{eq:modeseven} and \refer{eq:modesodd} are, however, difficult to solve exactly.

\section{A note on numerical methods}
\label{app:A}

The numerical computations were performed in the programming language \texttt{Julia} (version 1.10.3). The spatial domain was discretized on a uniform grid, typically using $n = 6000$ segments with lattice spacing $dx$. 

The second spatial derivative was approximated by a third-order central finite-difference stencil,
\begin{align}
\partial_x^2 \phi_i \approx \frac{1}{dx^2} \Bigg(
&\frac{1}{90}\phi_{i-3}
- \frac{3}{20}\phi_{i-2}
+ \frac{3}{2}\phi_{i-1} \nonumber \\
&- \frac{49}{18}\phi_i
+ \frac{3}{2}\phi_{i+1}
- \frac{3}{20}\phi_{i+2}
+ \frac{1}{90}\phi_{i+3}
\Bigg),
\end{align}
which significantly reduces numerical dispersion compared to the standard three-point scheme.

The field equation
\begin{align}
\partial_t^2 \phi
= \partial_x^2 \phi
- \frac{dU}{d\phi}
- \gamma(x)\,\partial_t \phi
\end{align}
was rewritten as a system of first-order equations at each lattice site,
\begin{align}
\partial_t \phi_i &= \pi_i, \\
\partial_t \pi_i &= \partial_x^2 \phi_i
- \frac{dU}{d\phi_i}
- \gamma_i \pi_i ,
\end{align}
where $\pi_i = \partial_t \phi_i$ and $\gamma_i$ is a position-dependent damping coefficient used near the boundaries to suppress spurious reflections. This yields a system of $2n$ coupled first-order ordinary differential equations (typically about $12\,000$ equations).

\subsection{Initial Conditions}

The initial boosted kink configuration was constructed by numerically integrating the BPS equation
\begin{align}
\frac{d\phi}{dx}
=
\frac{\sqrt{2V(\phi)}}{\sqrt{1 - v_{\mathrm{in}}^2}},
\end{align}
where $v_{\mathrm{in}}$ is the initial velocity. The static half-kink profile was obtained by solving this first-order equation with high precision and then mirrored to construct the full boosted kink profile,
\begin{align}
\phi_K(x)
=
\mathrm{sign}(x+p_0)\,
\phi_{\mathrm{half}}(|x+p_0|),
\end{align}
with $p_0$ denoting the initial position.

The kink--antikink collision initial condition was then formed as
\begin{align}
\phi(x,0)
&=
\phi_K(x;p_0)
-
\phi_K(x;-p_0)
- 1, \\
\partial_t \phi(x,0)
&=
- v_{\mathrm{in}} \phi_K'(x;p_0)
+ v_{\mathrm{in}} \phi_K'(x;-p_0),
\end{align}
ensuring two oppositely moving solitons centered symmetrically around the origin.

Due to the mirror symmetry of the setup, only half of the spatial domain was evolved numerically, reducing computational cost by a factor of two. The corresponding initial data were restricted to the right half of the lattice.

\subsection{Time Evolution}

The resulting initial-value problem was evolved using the \texttt{DifferentialEquations.jl} library. Time integration was performed with an explicit adaptive Runge--Kutta method, typically the Bogacki--Shampine 5/4 scheme (\texttt{BS5()}). In the construction of static profiles, the \texttt{Tsit5()} method was used for high-accuracy integration.

Overall, the procedure corresponds to a method-of-lines discretization: spatial derivatives are approximated on a fixed lattice, while time evolution is handled by a high-order adaptive ODE solver.


\end{document}